\documentclass[11pt,a4paper]{article}
\pdfoutput=1
\usepackage{jheppub}

\usepackage{url}

\allowdisplaybreaks[2]

\def\cA{\mathcal{A}}
\def\cB{\mathcal{B}}

\def\cD{\mathcal{D}}
\def\cE{\mathcal{E}}
\def\cF{\mathcal{F}}

\def\cO{\mathcal{O}}

\def\cU{\mathcal{U}}

\def\mint{\int_{-\infty}^\infty\!\cdots\!\int_{-\infty}^\infty}

\def\eK{\mathbb{K}}

\newcommand{\be}{\begin{equation}}
\newcommand{\ee}{\end{equation}}
\newcommand{\ba}{\begin{aligned}}
\newcommand{\ea}{\end{aligned}}

\def\Res{\mathop {\rm Res} \limits}

\DeclareMathOperator{\arccosh}{arccosh}

\def\ket#1{\left| #1 \right\rangle}

\def\({\left(}
\def\){\right)}

\def\wt#1{\widetilde{#1}}
\newcommand{\pd}{\partial}
\DeclareMathOperator{\real}{Re}
\DeclareMathOperator{\im}{Im}
\DeclareMathOperator{\Tr}{Tr}

\DeclareMathOperator{\vol}{vol}

\newcommand{\re}{{\rm e}}
\newcommand{\ri}{{\rm i}}
\newcommand{\rd}{{\rm d}}

\newcommand{\x}{\mathsf{x}}
\newcommand{\y}{\mathsf{y}}

\title{Calabi-Yau geometry and electrons on 2d lattices}

\author[a]{Yasuyuki Hatsuda,}
\author[b]{Yuji Sugimoto}
\author[c]{and Zhaojie Xu}

\affiliation[a]{D\'epartement de Physique Th\'eorique et Section de Math\'ematiques,\\
Universit\'e de Gen\`eve, Gen\`eve, CH-1211 Switzerland}
\affiliation[b]{Department of Physics, Graduate School of Science,\\
Osaka University, Toyonaka, Osaka 560-0043, Japan}
\affiliation[c]{Department of Physics and Astronomy,\\
Texas A\&M University, College Station, TX77843, USA}

\preprint{
\begin{flushright}
OU-HET-918
\end{flushright}
}

\abstract{
The B-model approach of topological string theory leads to difference equations
by quantizing algebraic mirror curves. It is known that these quantum mechanical systems
are solved by the refined topological strings.
Recently, it was pointed out that the quantum eigenvalue problem for a particular Calabi--Yau manifold, known as local $\mathbb{F}_0$,
is closely related to the Hofstadter problem for electrons on a two-dimensional square lattice.
In this paper, we generalize this idea to a more complicated Calabi--Yau manifold.
We find that the local $\cB_3$ geometry, which is a three-point blow-up of local $\mathbb{P}^2$, 
is associated with electrons on a triangular lattice. 
This correspondence allows us to use known results in condensed matter physics
to investigate the quantum geometry of the toric Calabi--Yau manifold.
}

\begin{document}

\maketitle

\renewcommand{\thefootnote}{\arabic{footnote}}
\setcounter{footnote}{0}
\setcounter{section}{0}

\section{Introduction}
Calabi--Yau (CY) manifolds play a central role in string theory.
Mirror symmetry relates two different CY manifolds in a highly non-trivial way.
In string theory, this symmetry is realized as a duality between type IIA string theory on
a CY manifold and type IIB on its mirror CY.
The same kind of duality is much more well-studied in the context of topological string theory,
a toy model of string theory.
There, mirror symmetry claims the equivalence of two different formulations of topological
string theory, called the A-model and the B-model.

Recently, there is interesting progress in the B-model formulation.
In the B-model, a toric CY geometry is characterized by an algebraic equation, called a mirror curve.
The generating function of the Gromov--Witten invariants of the CY manifold is constructed by this algebraic curve in principle.
Though the mirror curve is algebraic, it has an intriguing relation to quantum mechanical operators.
A relation between topological string theory and quantum mechanics turns back to the work \cite{ADKMV}.
The idea in \cite{ADKMV} was based on the fact that a brane in topological string theory plays the role of a wavefunction in
some quantum mechanical system.
This idea was recently realized more concretely in \cite{ACDKV}, based on the seminal work of Nekrasov and Shatashvili 
for supersymmetric gauge theories~\cite{NS}. 
A quantum mechanical system naturally appears by quantizing a mirror curve.
The remarkable conclusion in \cite{NS} is 
that the resulting quantum deformation of special geometry describes a special limit of the refined version of topological string theory,
where one of the two couplings of refined topological string theory is set to be zero.
Throughout this paper, we refer to this limit as the Nekrasov--Shatashvili limit or the NS limit for short.%
\footnote{In the spirit of \cite{ADKMV}, one also should be able to construct another quantum mechanical operator that
describes the standard (or unrefined) topological string theory from the mirror curve.
An attempt to this issue is found in \cite{GS} for example.
In this approach, the corresponding quantum operators can be written in closed form for a few mirror curves with genus zero,
but for higher genus mirror curves, they receive complicated quantum corrections in general.  
The quantization scheme here is much simpler than the one in \cite{GS}.
One can easily write down a quantum operator for a given mirror curve with genus greater than one.
Nevertheless, one can also extract results for the unrefined topological strings from the obtained simple quantum operator, as explained below.}

The correspondence between topological string theory and quantum mechanics
was further clarified in \cite{GHM1}.
The work \cite{GHM1} has two important aspects.
On one side of the story, the correspondence allows us to solve the quantum eigenvalue problem by using the result in topological string theory.
In fact, an exact form of the so-called spectral determinant was conjectured in \cite{GHM1} in terms of the refined topological invariants
of the corresponding CY manifold.
The conjecture is nonperturbatively valid for any Planck constant. The eigenvalues were then read off as zeros of this spectral determinant.
The zero locus condition led to a kind of quantization conditions for the eigenvalues.
Though the quantization condition originally derived in \cite{GHM1} looked complicated, it was pointed out in \cite{WZH}
that this condition is rewritten as a much simpler form.
Such quantization conditions were confirmed to give the correct eigenvalues for a class of integrable systems associated with
toric CY manifolds \cite{HM, FHM}.

On the other side of the story, the correspondence provides a nonperturbative result on topological string theory
from the quantum mechanical operator.
One of the main claims in \cite{GHM1} is that the spectral determinant describes the unrefined topological string free energy 
(not the refined free energy in the NS limit) in a 't Hooft-like strong coupling limit.
Since the spectral determinant is constructed nonperturbatively in the Planck constant,
the unrefined topological string free energy receives non-trivial corrections in this 't Hooft-like limit.
For some particular toric CY manifolds, this approach naturally leads to a new matrix model description of (unrefined) topological strings \cite{MZ, KMZ}.
Such matrix models are expected to give a nonperturbative realization of topological string theory beyond the perturbative definition.
The validity of the conjectural spectral determinant in \cite{GHM1} was confirmed for many examples \cite{GKMR, CGM, BGT, Grassi, SWH, CGuM}.
The exact eigenfunctions were also conjectured in \cite{MZ2} along this approach.
Furthermore, very recently, the matrix model description for local $\mathbb{P}^2$ was compared with another nonperturbative approach 
based on resurgence theory \cite{CSMS}. Both results show remarkable agreement.

\begin{figure}[t]
\begin{center}
  \begin{minipage}[b]{0.4\linewidth}
    \centering
    \includegraphics[width=4cm]{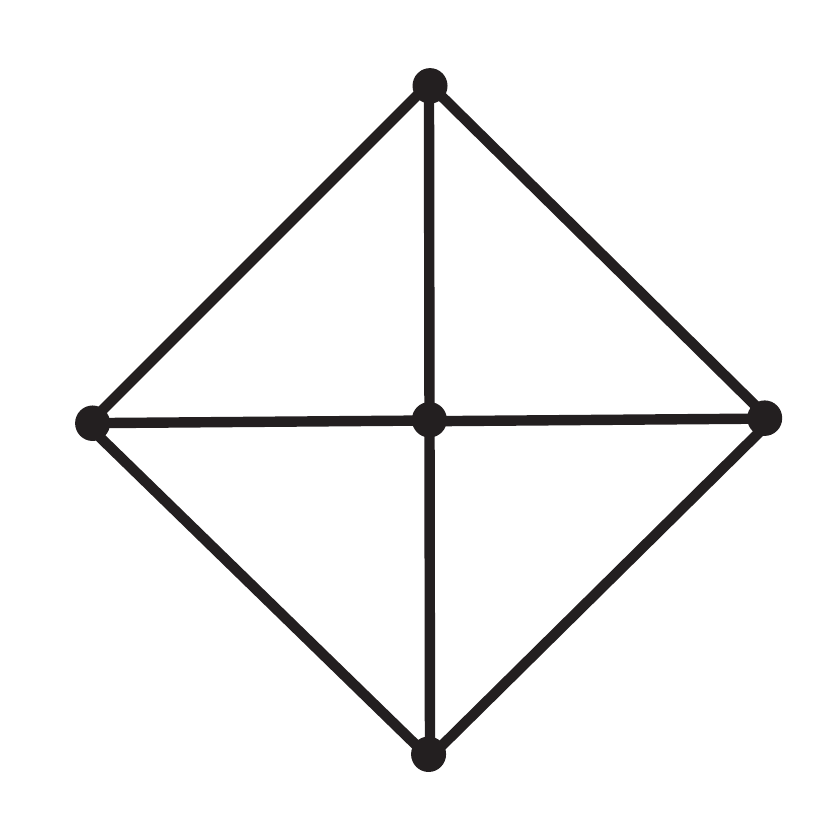}
  \end{minipage} \hspace{1cm}
  \begin{minipage}[b]{0.4\linewidth}
    \centering
    \includegraphics[width=6cm]{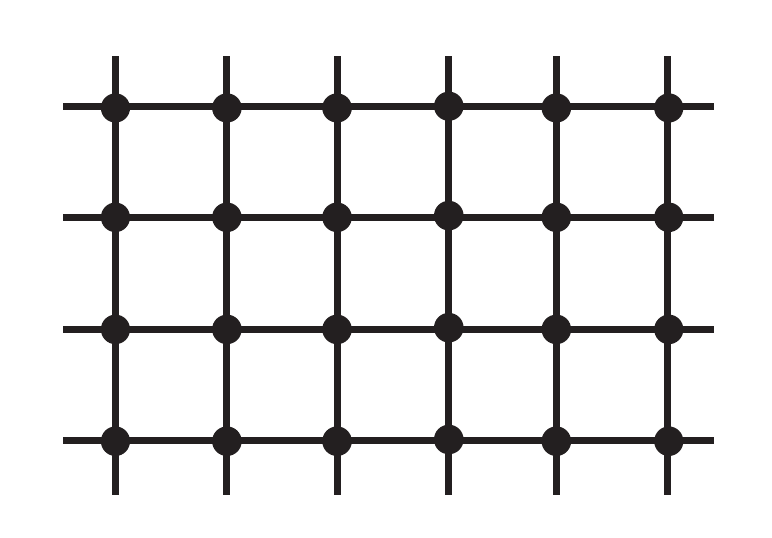}
  \end{minipage} 
\end{center}
  \caption{Left: The toric diagram of local $\mathbb{F}_0$. Right: The two-dimensional lattice corresponding to the Hofstadter problem.
An electron on the lattice can hop to any nearest neighbors.}
  \label{fig:F0}
\end{figure}

However, this is not the end of the story. 
The relation between Calabi--Yau geometries and quantum mechanics sheds light on a new connection with condensed matter physics.
Recently, it was also found in \cite{HKT} that the quantum mechanical system associated with
the CY threefold, called local $\mathbb{F}_0$, is closely related to the Hofstadter problem \cite{Hof} in electrons on a two-dimensional lattice
with a perpendicular magnetic field.
To make the explanation clearer, let us briefly review the idea in \cite{HKT}.
The toric diagram of local $\mathbb{F}_0$ is shown in the left of Figure~\ref{fig:F0}.
The resulting Hamiltonian of the quantum system associated with this geometry is the following difference operator
\be
H_{\mathbb{F}_0}=\re^{\x}+\re^{-\x}+\re^{\y}+\re^{-\y},\qquad [\x , \y]=\ri \hbar.
\label{eq:H-F0}
\ee
As proved in \cite{KM}, the inverse operator $H_{\mathbb{F}_0}^{-1}$ is in the trace class on the Hilbert space $L^2(\mathbb{R})$. 
Therefore the eigenvalue problem $H_{\mathbb{F}_0} \psi(x)=\cE \psi(x)$ for $\psi(x) \in L^2(\mathbb{R})$ has an infinite number of \textit{discrete} eigenvalues
$\cE_n$ ($n=0,1,2,\dots$).  

On the other hand, the Hofstadter problem is an eigenvalue problem for electrons on the 2d lattice shown in the right of Figure~\ref{fig:F0},
in the presence of a magnetic flux perpendicular to the lattice plane.
In the tight-binding approximation, its Hamiltonian is given by
\be
H_\text{Hofstadter}=T_x+T_x^\dagger+T_y+T_y^\dagger,\qquad  T_x T_y=\re^{\ri \phi} T_y T_x,
\label{eq:Hof}
\ee
where $T_x$ and $T_y$ are magnetic translation operators in each direction, 
and $\phi$ is a magnetic flux through an elementary plaquette.
It was shown in \cite{Hof} that, for $\phi=2\pi a/b$ with coprime integers $a$ and $b$, 
the spectrum of \eqref{eq:Hof} has the $b$-subband structure and that its shape as a function of $\phi$ is fractal.
This is well-known as the Hofstadter butterfly.

One can see that the above two Hamiltonians are very similar, but there is a big difference.
The latter is periodic in $x$- and $y$-directions since the electrons are put  on the 2d lattice. 
The former has no such periodic structure on the real lines of $x$ and $y$.
This suggests us to perform an analytic continuation $\x \to \ri \x$ and $\y \to \ri \y$ in \eqref{eq:H-F0}.
Note that the similar Hamiltonian, in which only one of the two variables is analytically continued, was recently studied in \cite{Krefl3}.
In our perspective, it is more natural to do the analytic continuation of both $\x$ and $\y$.
The exponentiated operators satisfy $\re^{\ri \x}\re^{\ri \y}=\re^{-\ri \hbar} \re^{\ri \y}\re^{\ri \x}$.
Now it is clear that the analytic continued Hamiltonian of \eqref{eq:H-F0} and the Hofstadter Hamiltonian \eqref{eq:Hof} are identical under 
\be
T_x \leftrightarrow \re^{\ri \x}, \qquad T_y \leftrightarrow \re^{\ri \y},\qquad \phi \leftrightarrow -\hbar.
\label{eq:identification}
\ee
This means that the magnetic flux $\phi$ corresponds to the Planck constant $\hbar$ in the quantum system of
the toric CY.
This similarity implies that the two eigenvalue problems for \eqref{eq:H-F0} and for \eqref{eq:Hof}
are closely related. However, it is not so obvious to find the precise relation between them.
It was found in \cite{HKT} that the information on the spectrum of the Hofstadter problem is encoded
in the so-called the quantum corrected period around a cycle in the mirror geometry of local $\mathbb{F}_0$.%
\footnote{It is known that the spectral problem of \eqref{eq:H-F0} is solved in terms of the quantum period around the other cycle.
We will review it in the case of local $\cB_3$ in the next section.}
More precisely, it was shown that the branch cut structure of the (normalizable)  K\"ahler modulus in quantum geometry of local $\mathbb{F}_0$
is identical to the band spectrum of the Hofstadter problem.

\begin{figure}[t]
\begin{center}
  \begin{minipage}[b]{0.4\linewidth}
    \centering
    \includegraphics[width=4cm]{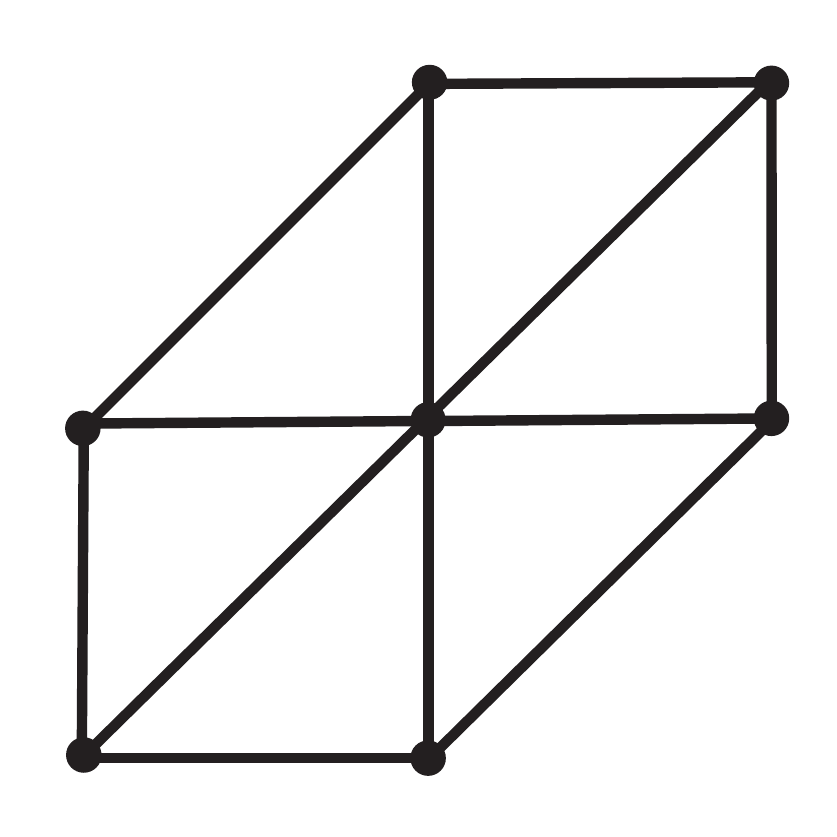}
  \end{minipage} \hspace{1cm}
  \begin{minipage}[b]{0.4\linewidth}
    \centering
    \includegraphics[width=6cm]{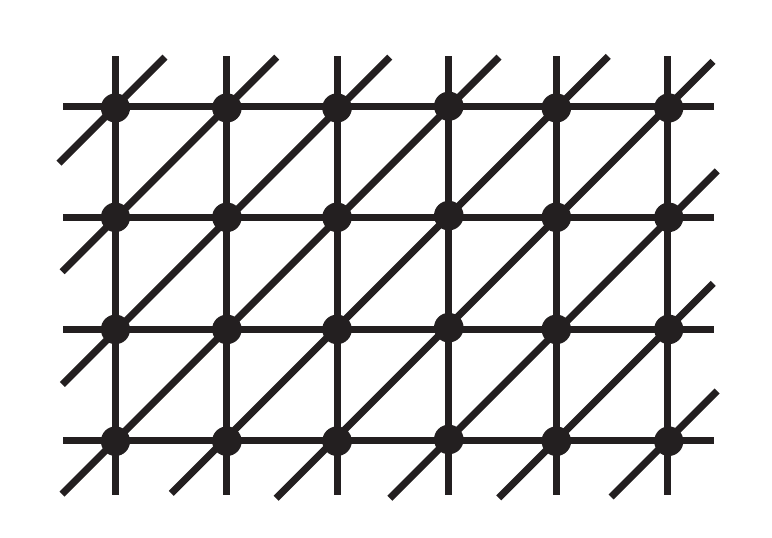}
  \end{minipage} 
\end{center}
  \caption{Left: The toric diagram of local $\cB_3$. Right: The two-dimensional triangular lattice.
An electron on the lattice can hop to any nearest neighbors as well as to two next-nearest neighbors connected by the solid lines. One can see that
the toric diagram of local $\cB_3$ is embedded in this lattice system.}
  \label{fig:B3}
\end{figure}

It is natural to ask whether this correspondence holds for other examples.
This is indeed the case.
In this paper, we particularly consider another relevant toric CY manifold, known as local $\cB_3$.
The local $\cB_3$ geometry is a three-point blow-up of local $\mathbb{P}^2$ at generic points, and
its toric diagram is given by the left of Figure~\ref{fig:B3}.
As will be reviewed in the next section, the Hamiltonian for this geometry is given by
\be
H_{\cB_3}=\re^{\x}+\re^{-\x}+\re^{\y}+\re^{-\y}+\re^{\x+\y}+\re^{-\x-\y}.
\ee
The counterpart to this Hamiltonian is easily found in the literature.
It is just the Hamiltonian for electrons on a triangular lattice, shown in the right of Figure~\ref{fig:B3}.
Its explicit form takes the following form
\be
H_\text{tri}
=T_x+T_x^\dagger+T_y+T_y^\dagger+ \re^{-\ri \phi/2} T_x T_y+\re^{\ri \phi/2} T_y^\dagger T_x^\dagger.
\ee
The spectrum of this Hamiltonian was studied long time ago \cite{CW}.
The problem is also solved by the Bethe ansatz method \cite{WZ, FK}.
The main purpose of this paper is to study the correspondence between these two Hamiltonians in great detail.
We confirm that these two spectral problems are actually interrelated.
Since the spectral problem for the triangular lattice has been well-studied, we can use known results about it
in order to investigate the spectral problem for local $\cB_3$.
Conversely, the quantum local $\cB_3$ geometry knows the complete spectral information
for the electrons on the triangular lattice.

This paper is organized as follows. In the next section, we will review the conjectural solution
to the eigenvalue problem associated with toric CY  manifolds.
We focus on the case of local $\cB_3$.
In section~\ref{sec:CY-electrons}, we will see a concrete connection with the 2d electron system on the triangular lattice.
As in \cite{HKT}, we will identify the branch cuts of the K\"ahler modulus of quantum geometry in local $\cB_3$
with the band spectrum of the Hamiltonian for the triangular lattice.
Moreover, the imaginary part of the K\"ahler modulus is identified with the density of states. 
In section~\ref{sec:conclusion}, some remarks are given. In particular, we will raise two open problems.
In appendices, we collect some useful results on the eigenvalue problem for local $\cB_3$.

\section{Quantum eigenvalue problem in local $\cB_3$ geometry}
In this section, we start by reviewing an exact solution to a quantum eigenvalue problem
associated with a toric Calabi--Yau threefold.
In \cite{GHM1}, it was conjectured that this eigenvalue problem is exactly solved by using
the refined topological string results. This conjecture is now confirmed for many examples.
We are particularly interested in local $\cB_3$, but there seems to be no explicit result on this geometry in the literature.
Therefore, as a review, we here explain how to get the exact eigenvalues for this geometry in detail.
This is a straightforward application of the previous works \cite{HW, GHM1, GKMR} to the local $\cB_3$ geometry.

Our starting point is the B-model description of CYs.
The local $\cB_3$ geometry is a three-point blow-up of the local $\mathbb{P}^2$ geometry.
It is known that its mirror geometry is described by an algebraic equation called a mirror curve. 
The toric diagram of local $\cB_3$ is shown in the left of Figure~\ref{fig:B3}.
The mirror curve of local $\cB_3$ takes the form
\be
\re^{x}+\re^y+\re^{-x-y}+m_1 \re^{-x}+m_2 \re^{-y}+m_3 \re^{x+y}=\cE,
\label{eq:mirror-B3}
\ee
where $(m_1,m_2,m_3)$ are ``mass parameters,'' and $\cE$ is the ``true modulus'' in the terminology of \cite{HKP}.%
\footnote{Sometimes these two are also called non-normalizable moduli and normalizable moduli, respectively.}
The mass parameters are given by hand, while the true modulus plays the role of the energy in the quantum mechanical system.
These four are actually complex structure parameters of the mirror CY of local $\cB_3$.
If we turn off the mass parameters $(m_1,m_2, m_3) \to (m_1, m_2, 0) \to (m_1, 0, 0) \to (0,0,0)$ in turn,
the geometry is reduced to local $\cB_2$, local $\mathbb{F}_1$ and local $\mathbb{P}^2$, respectively.
The spectral problems for these reduced geometries were analyzed in \cite{GHM1, GKMR}.%
\footnote{As we will see later, the instanton expansion of the prepotential (or the free energy) is
singular in the naive blow-down limit $m_i \to 0$. However, if setting $m_i=0$ in the mirror curve \eqref{eq:mirror-B3} and repeating the same computation below,
one reproduces the results in \cite{GHM1, GKMR}. This implies that the large radius limit $t \to \infty$ and the blow-down limit $m_i \to 0$ do not commute
in general.}

The quantization of the mirror curve leads to a difference equation.
As usual, we impose the commutation relation
\be
[\x,\y]=\ri \hbar.
\ee
To avoid ambiguity of operator ordering, we need the quantization procedure.
This is not unique in general.
Here, we follow the proposal in \cite{GHM1}.
We take the following quantization method
\be
\re^{ax+by} \to \re^{a  \x+ b \y}=q^{-ab/2}\re^{a \x} \re^{b \y},
\label{eq:quantization}
\ee
where $q=\re^{\ri \hbar}$.
Then, the quantum Hamiltonian of the mirror curve \eqref{eq:mirror-B3} is simply given by
\be
H=\re^{\x}+\re^{\y}+\re^{-\x-\y}+m_1 \re^{-\x}+m_2 \re^{-\y}+m_3 \re^{\x+\y}.
\label{eq:H-B3}
\ee
Note that this Hamiltonian is automatically a self-adjoint operator due to the quantization scheme \eqref{eq:quantization}.
Since $\re^{\pm \y}=\re^{\mp \ri \hbar \pd_x}$ are difference operators,
the quantum eigenvalue problem $H \psi(x)=\cE \psi(x)$ leads to the following difference equation
\be
\ba
&\re^{x} \psi(x)+\psi(x-\ri \hbar)
+q^{-1/2}\re^{-x} \psi(x+\ri \hbar)\\
&\qquad+m_1 \re^{-x} \psi(x)+m_2 \psi(x+\ri \hbar)+m_3 q^{-1/2} \re^{x} \psi(x-\ri \hbar)
=\cE \psi(x),
\ea
\label{eq:diff-eq}
\ee
As explained in \cite{GHM1, KM}, if we require the square integrability and the analyticity in the strip $| \im x | \leq \hbar$ 
to the wave function $\psi(x)$,
the difference equation has solutions only for an infinite number of discrete values of $\cE$.
The main consequence in \cite{GHM1, WZH} is that all of these eigenvalues are exactly fixed by a quantization condition
including all the quantum corrections. 
We will see it in the following two subsections

The quantization scheme \eqref{eq:quantization} is very simple. 
One can write down a quantum operator for a given mirror curve.
As we will see later, this simple scheme describes
the refined topological strings in the NS limit, semiclassically.
Another quantization scheme is also found, for example, in \cite{GS}.
Though the quantization in \cite{GS} is not in general easy to find a quantum operator for a given curve in closed form,
the semiclassical expansion of the quantum mechanical system describes the genus expansion of the unrefined topological string.
Surprisingly, it was shown in \cite{GHM1} that the quantization scheme here also describes the \textit{unrefined}
topological strings in the \textit{strong coupling} regime of the quantum system.
We will review this fact in appendix~\ref{sec:spectral}.

\subsection{Classical analysis: Bohr--Sommerfeld quantization condition}\label{subsec:classical}
\begin{figure}[t]
\begin{center}
  \begin{minipage}[b]{0.4\linewidth}
    \centering
    \includegraphics[width=5.5cm]{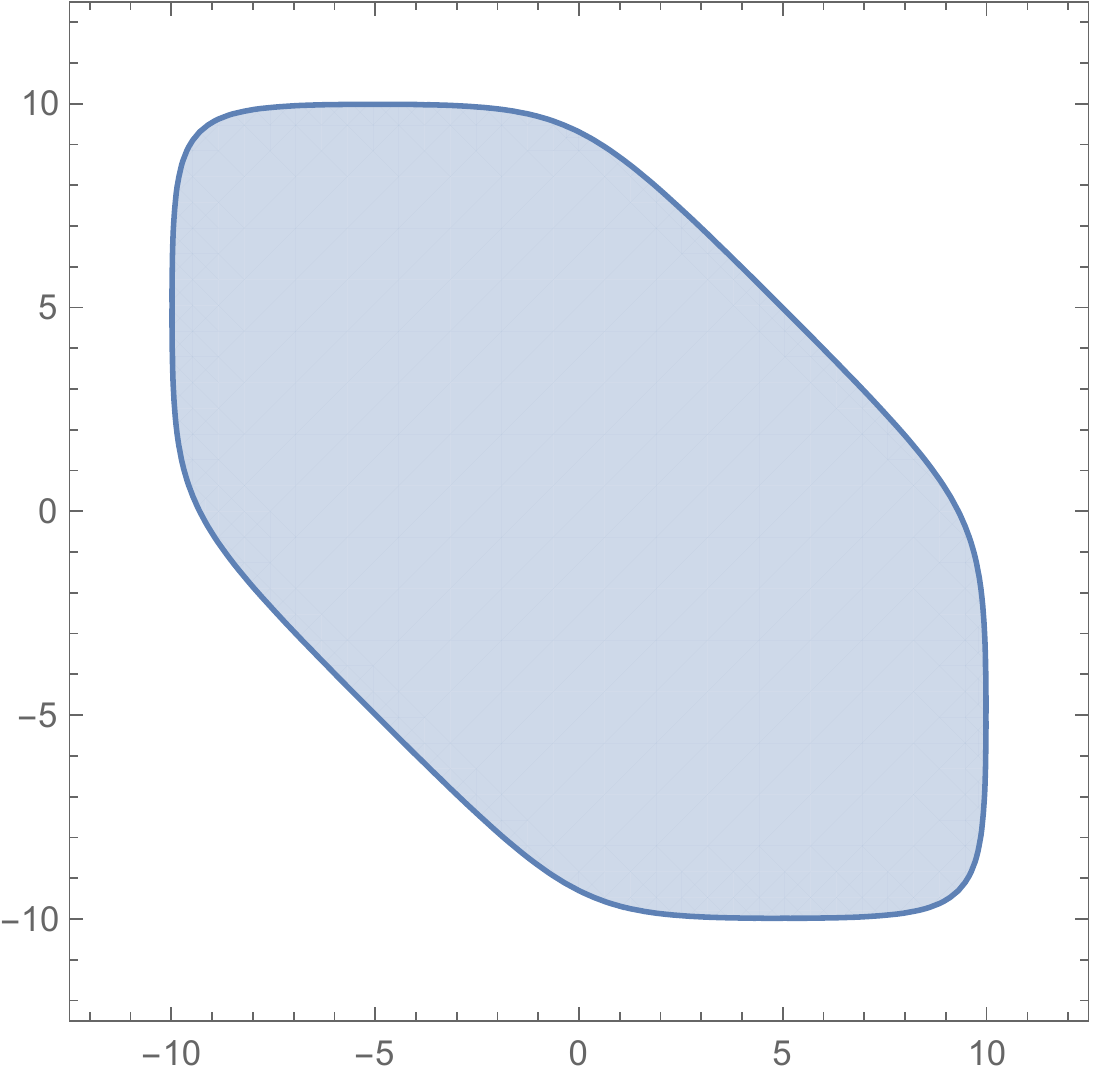} \\
   (a)
  \end{minipage} \hspace{1cm}
  \begin{minipage}[b]{0.4\linewidth}
    \centering
    \includegraphics[width=5.5cm]{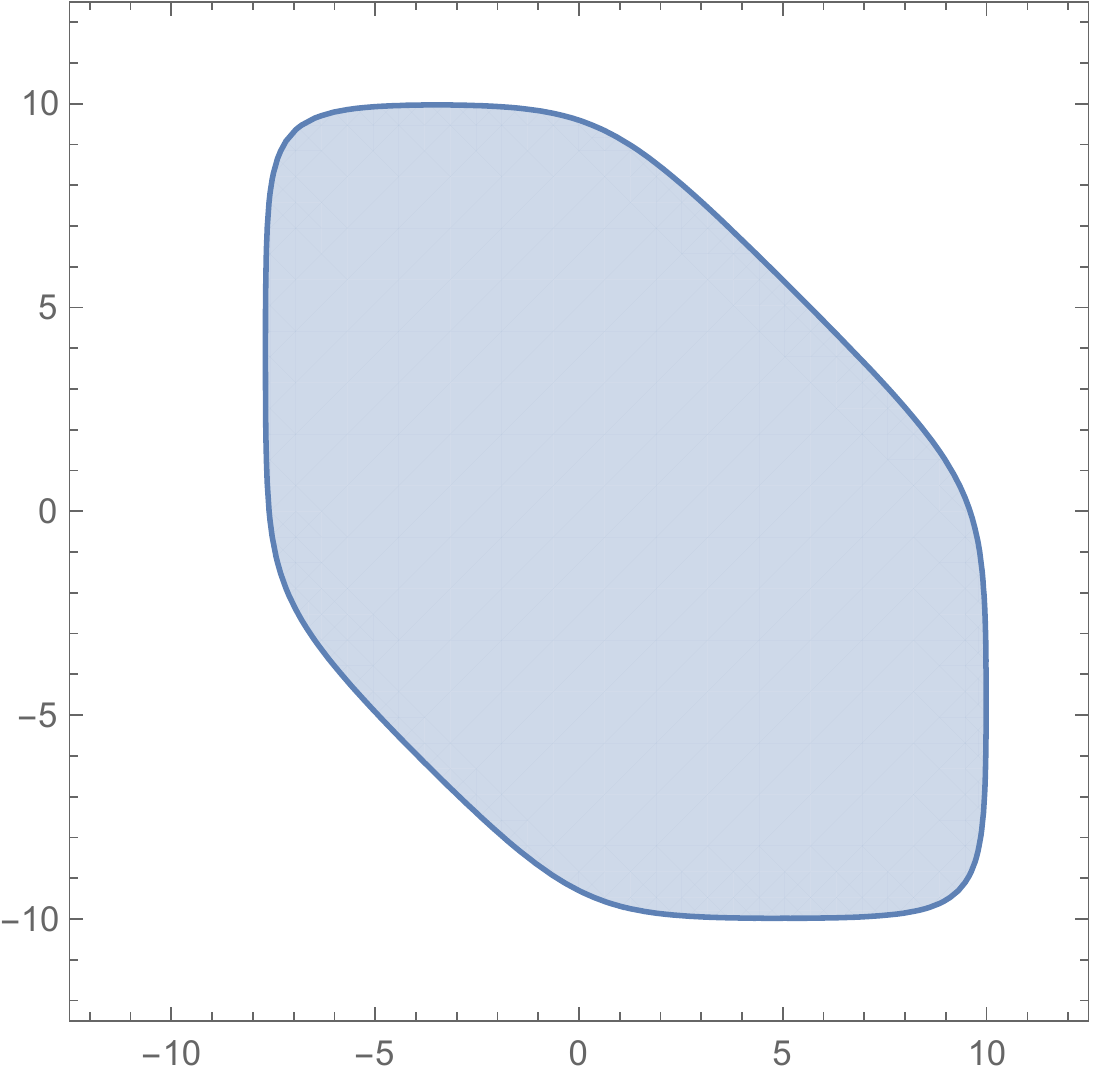} \\
    (b)
  \end{minipage} 
\end{center}
  \caption{We show the region $\re^{x}+\re^{y}+\re^{-x-y}+m_1\re^{-x}+m_2 \re^{-y}+m_3 \re^{x+y} \leq \cE$ in the phase plane $(x,y)$. 
The energy is set to be $\log \cE=10$, and the mass parameters are (a) $(m_1,m_2,m_3)=(1,1,1)$ and (b) $(m_1,m_2,m_3)=(10,1,1/2)$, respectively.}
  \label{fig:Fermi}
\end{figure}

In the semiclassical limit $\hbar \to 0$, the quantization condition is described by the well-known Bohr--Sommerfeld condition:
\be
\oint_B \! \rd x\,  y(x, \cE)=2\pi \hbar \( n+\frac{1}{2} \).
\label{eq:BS}
\ee
where the integration path $B$ on the left hand side is a cycle going around two turning points. 
Geometrically, this integral gives the area of the region surrounded by the curve \eqref{eq:mirror-B3}
in the phase plane $(x,y)$:
\be
\vol_0(\cE):=\oint_B \! \rd x\,  y(x, \cE)=\int_{x_1(\cE)}^{x_2(\cE)} \rd x\, (y_+(x, \cE)-y_-(x, \cE) ),
\label{eq:vol0-int}
\ee
where $y_\pm(x,\cE)$ are two branches determined by the equation \eqref{eq:mirror-B3},
and $x_1(\cE)$ and $x_2(\cE)$ are the two turning points (i.e., $y_+(x_i(\cE),\cE)=y_-(x_i(\cE), \cE)$ for $i=1,2$).
See Figure~\ref{fig:Fermi}.
It is not easy to evaluate this integral by a brute-force method since $y_\pm(x, \cE)$ are very complicated.
To evaluate it, we can use results in special geometry.

It is known that special geometry relates the integral of $\vol_0(\cE)$,
which is usually called the B-period in the literature, to the prepotential (or the genus zero free energy)
of the corresponding CY manifold:
\be
\frac{\pd F_0(t)}{\pd t}=\vol_0 (\cE)=\oint_B \rd x\,  y(x, \cE).
\label{eq:cB}
\ee
Since the mirror curve \eqref{eq:mirror-B3} defines a genus one Riemann surface,
one can consider another cycle, usually called the A-cycle.
The K\"ahler modulus $t$ in the above equation is just related to the energy $\cE$ by this A-period:
\be
t=\oint_A \rd x \, y(x, \cE).
\label{eq:cA}
\ee
These two equations determine the prepotential as a function of $t$.
At the technical level, there is an efficient way to compute these periods exactly.
As explained in \cite{HKP}, 
the computation of these periods is mapped to the periods for the Weierstrass normal form of the elliptic curve:
\be
Y^2=4X^3-g_2 X-g_3.
\label{eq:elliptic-curve}
\ee
To go from the mirror curve \eqref{eq:mirror-B3} to the Weierstrass form \eqref{eq:elliptic-curve},
so-called  Nagell's algorithm is used. See \cite{HKP} in detail.
Fortunately, in our case, we can directly use the result (A.22) in \cite{HKP} without any new calculations, 
and the coefficients $g_2$ and $g_3$ are explicitly given by
\be
\ba
g_2&=\frac{1}{12z^4}[ 1-8(m_1+m_2+m_3)z^2-24(1+m_1 m_2 m_3)z^3 \\
&\quad+16(m_1^2+m_2^2+m_3^2-m_1 m_2-m_2 m_3-m_3 m_1)z^4], \\
g_3&=\frac{1}{216z^6}[ 1-12(m_1+m_2+m_3)z^2-36(1+m_1 m_2 m_3)z^3 \\
&\quad+24(2m_1^2+2m_2^2+2m_3^2+m_1 m_2+m_2 m_3+m_3 m_1)z^4 \\
&\quad+144(m_1+m_2+m_3)(1+m_1 m_2 m_3)z^5 \\
&\quad+8(-8m_1^3-8m_2^3-8m_3^3+12m_1^2 m_2+12m_2^2 m_3+12m_3^2 m_1\\
&\quad+12m_1 m_2^2+12m_2 m_3^2+12m_3 m_1^2+27+6m_1 m_2 m_3+27m_1^2 m_2^2 m_3^2)z^6],
\ea
\label{eq:g2g3}
\ee
where $z=1/\cE$.
Then, the periods are written in closed forms
\be
\ba
\frac{\pd t}{\pd z}&=-\frac{1}{2\pi z^2} \frac{2}{\sqrt{e_1-e_3}} \eK \( \frac{e_2-e_3}{e_1-e_3} \), \\
\frac{\pd^2 F_0}{\pd z \pd t}&=-\frac{1}{z^2} \frac{2}{\sqrt{e_1-e_3}} \eK \( \frac{e_1-e_2}{e_1-e_3} \).
\ea
\label{eq:classical-periods}
\ee
where $\mathbb{K}(m)$ is the complete elliptic integral of the first kind, 
and $e_1$, $e_2$ and $e_3$ are three roots of the elliptic curve \eqref{eq:elliptic-curve}.
We have to choose them to reproduce the correct asymptotics of the A- and B-periods in $z \to 0$.
The A-period has a logarithmic divergence, while the B-period has a double logarithmic divergence in this limit.
The explicit forms of $(e_1, e_2, e_3)$ are very complicated, but we can fix them by the behavior in the limit $z \to 0$ as follows:
\be
\ba
e_1&=\frac{1}{6z^2}-\frac{2}{3}(m_1+m_2+m_3)-2(1+m_1 m_2 m_3)z+\cO(z^2) ,\\
e_2&=-\frac{1}{12z^2}+\frac{1}{3}(m_1+m_2+m_3)+(1+m_1m_2m_3+2\sqrt{m_1 m_2 m_3})z+\cO(z^2) ,\\
e_3&=-\frac{1}{12z^2}+\frac{1}{3}(m_1+m_2+m_3)+(1+m_1m_2m_3-2\sqrt{m_1 m_2 m_3})z+\cO(z^2) .
\ea
\ee 
Plugging these expansions into the first equation in \eqref{eq:classical-periods}, one finds
\be
\ba
-t&=\log z+(m_1+m_2+m_3)z^2+2(1+m_1 m_2 m_3)z^3 \\
&\quad+\frac{3}{2}(m_1^2+m_2^2+m_3^2+4m_1 m_2+4m_2 m_3+4m_3 m_1)z^4+\cO(z^5),
\ea
\label{eq:A-period}
\ee
where we fixed the integration constant so that $Q=\re^{-t}=z+\cO(z^2)$ in $z\to 0$.
Inverting this, one gets the so-called mirror map
\be
\ba
z&=Q[1-(m_1+m_2+m_3)Q^2-2(1+m_1 m_2 m_3)Q^3\\
&\quad+(m_1^2+m_2^2+m_3^2-m_1m_2-m_2m_3-m_3m_1)Q^4+\cO(Q^5)].
\ea
\label{eq:mirror-map}
\ee
Similarly, from the second equation in \eqref{eq:classical-periods}, one obtains
\be
\ba
\frac{\pd F_0}{\pd t}&=3\log^2 z+\log(m_1 m_2 m_3) \log z+C_0 \\
&\quad+\(\frac{1}{m_1}+\frac{1}{m_2}+\frac{1}{m_3}+m_1 m_2+m_2 m_3+m_3 m_1 \)z \\
&\quad+\biggl[ (m_1+m_2+m_3)(\log (m_1 m_2 m_3)+6\log z)
+4(m_1+m_2+m_3)\\
&\quad-\frac{1}{4m_1^2}-\frac{1}{4m_2^2}-\frac{1}{4m_3^2}
-\frac{m_1^2 m_2^2}{4}-\frac{m_2^2 m_3^2}{4}-\frac{m_3^2 m_1^2}{4}\biggr]z^2
+\cO(z^3).
\ea
\label{eq:dF0}
\ee
The integration constant $C_0$ is not fixed in this way. We fix it by comparing to $\vol_0(\cE)$.
From the numerical experiment, we observe the following asymptotic behavior of $\vol_0(\cE)$ in $\cE \to \infty$:
\be
\vol_0(\cE)=3\log^2 \cE-\log(m_1 m_2 m_3)\log \cE-\pi^2-\frac{1}{2}(\log^2 m_1+\log^2 m_2+\log^2 m_3 )
+\cO(\cE^{-1}).
\ee
Comparing this expansion with \eqref{eq:dF0}, the integration constant should be fixed by
\be
C_0=-\pi^2-\frac{1}{2}(\log^2 m_1+\log^2 m_2+\log^2 m_3 ).
\label{eq:C0}
\ee
Substituting the mirror map \eqref{eq:mirror-map} into \eqref{eq:dF0},
we finally obtain the prepotential
\be
\ba
F_0(t)&=t^3-\frac{\log(m_1 m_2 m_3)}{2}t^2+C_0 t+\wt{C}_0+F_0^\text{inst}(t), \\
F_0^\text{inst}(t)&=-\(\frac{1}{m_1}+\frac{1}{m_2}+\frac{1}{m_3}+m_1 m_2+m_2 m_3+m_3 m_1 \)\re^{-t}\\
&\quad -\frac{1}{8}\biggl( 16(m_1+m_2+m_3)-\frac{1}{m_1^2}-\frac{1}{m_2^2}-\frac{1}{m_3^2} \\
&\quad -m_1^2 m_2^2-m_2^2 m_3^2-m_3^2 m_1^2 \biggr)\re^{-2t}+\cO(\re^{-3t}). 
\ea
\label{eq:F0}
\ee
Another integration constant $\wt{C}_0$ is not relevant in our analysis.

Let us remark on the blow-down limit.
It is obvious that the A-period \eqref{eq:A-period} admits the blow-down limit $m_i \to 0$.
One can confirm that this limit correctly reproduces the A-periods for local $\mathbb{P}^2, \mathbb{F}_1, \cB_2$.
See  \cite{GKMR} for instance.
On the other hand, the instanton expansion \eqref{eq:F0} of the prepotential is not well-defined in the limit $m_i \to 0$.
However, if setting $m_i \to 0$ from the beginning \eqref{eq:mirror-B3}, one can obtain the correct prepotential.
Let us consider the case of local $\cB_2$ ($m_3=0$).
After doing the same computation above, we obtain
\be
\ba
F_0^{\cB_2}(t)&=\frac{7t^3}{6}-\frac{\log(m_1 m_2)}{2}t^2+C_0^{\cB_2} t+\wt{C}_0^{\cB_2}
-\( \frac{1}{m_1}+\frac{1}{m_2}+m_1 m_2\) \re^{-t} \\
&\quad-\frac{1}{8}\( 16m_1+16m_2-\frac{1}{m_1^2}-\frac{1}{m_2^2}+m_1^2 m_2^2 \)\re^{-2t}+\cO(\re^{-3t}).
\ea
\ee
This is indeed the prepotential for local $\cB_2$ (see (4.88) in \cite{GKMR}).
It is not so easy to reproduce this result only from the large radius expansion \eqref{eq:F0} for local $\cB_3$.

In particular, for $m_1=m_2=m_3=1$, things are much simpler.
In this case, by shifting the $y$-variable by $y \to y-x/2$, the mirror curve becomes
\be
2\cosh x+4\cosh \frac{x}{2} \cosh y=\cE.
\ee
Then the periods are written as
\be
\ba
t&=\frac{1}{2\pi \ri} \int_{x_-}^{x_+} \rd x \, \arccosh \( \frac{\cE-2\cosh x}{4\cosh \frac{x}{2} } \), \\
\frac{\pd F_0}{\pd t}&=4 \int_{0}^{x_-} \rd x \, \arccosh \( \frac{\cE-2\cosh x}{4\cosh \frac{x}{2} } \),
\ea
\ee
where $x_\pm$ are defined for $\cE>6$ by
\be
\frac{\cE-2\cosh x_{\pm}}{4\cosh \frac{x_\pm}{2}}=\mp 1,\qquad x_\pm>0.
\ee
The derivative of these integral with respect to $\cE$ can be evaluated exactly.
We find the following expressions:
\be
\ba
\frac{\pd t}{\pd \cE}&= \frac{2}{\pi \sqrt{\cE^2-12+8\sqrt{\cE+3}}}
\eK \( \frac{16\sqrt{\cE+3}}{\cE^2-12+8\sqrt{\cE+3}} \),\\
\frac{\pd^2 F_0}{\pd \cE \pd t}&= \frac{8}{\sqrt{\cE^2-12+8\sqrt{\cE+3}}}
\eK \( \frac{\cE^2-12-8\sqrt{\cE+3}}{\cE^2-12+8\sqrt{\cE+3}} \)
\ea
\label{eq:classical-periods-2}
\ee
From these, one immediately finds
\be
-t=\log z+3z^2+4z^3+\frac{45}{2}z^4+72z^5+340z^6+\cO(z^7).
\label{eq:t-cl}
\ee
and
\be
\ba
F_0(t)&=t^3-\pi^2 t+\wt{C}_0+F_0^\text{inst}(t), \\
F_0^\text{inst}(t)&=-6\re^{-t}-\frac{21}{4}\re^{-2t}-\frac{56}{9}\re^{-3t}-\frac{405}{32}\re^{-4t}-\frac{3756}{125}\re^{-5t}
-\frac{751}{9}\re^{-6t}+\cO(\re^{-7t}).
\label{eq:F0-equal-mass}
\ea
\ee
One can check that the functions
\be
w_A(z):=\frac{\pd t}{\pd z},\qquad w_B(z):= \frac{\pd^2 F_0}{\pd z \pd t},
\ee
both satisfy the following second order differential equation
\be
\ba
z^2(1+2z)(1+3z)(1-6z) w_i''(z)+z(3-4z-120z^2-216z^3)w_i'(z) \\
+(1-2z-96z^2-216z^3)w_i(z)=0,\qquad i=A, B.
\ea
\label{eq:PF}
\ee
This differential equation can be regarded as a Picard--Fuchs equation.
Interestingly, we notice that the same differential equation appears in
the so-called mass deformed $E_8$ del Pezzo geometry with three non-vanishing masses studied in \cite{GKMR}.
We will see this relation in more detail in appendix~\ref{sec:E8}.

Let us summarize the computation here. Using the special geometry relation, the Bohr--Sommerfeld quantization condition \eqref{eq:BS}
is rewritten in terms of the prepotential of the corresponding CY geometry:
\be
\frac{\pd F_0(t)}{\pd t}=2\pi \hbar \( n+\frac{1}{2} \).
\label{eq:BS-2}
\ee
This fixes the K\"ahler modulus $t$ for a given non-negative integer $n$.
Once this value is known, the energy $\cE$ is recovered by the mirror map \eqref{eq:mirror-map}.
However, one has to recall that the Bohr--Sommerfeld quantization condition is
the first order semiclassical approximation. The eigenvalues obtained in this way do not give the exact answer.
In the next subsection, we will review how to incorporate the quantum corrections to the Bohr--Sommerfeld quantization condition.

\subsection{Quantum corrections: exact quantization condition}\label{subsec:quantum}
On the quantum corrections to the Bohr--Sommerfeld quantization condition,
we can use the old work of Dunham \cite{Dunham}.
Though his argument is for the Schr\"odinger equation,
the final result is applicable to our setup.
Let us consider a WKB-type formal solution:
\be
\ba
\psi(x)=\exp \left[ \frac{\ri}{\hbar} \int^x \rd x' \, Y(x', \cE, \hbar) \right],\qquad
Y(x, \cE, \hbar)=\sum_{\ell=0}^\infty \hbar^\ell y_\ell(x, \cE).
\ea
\ee
As in the standard WKB analysis,
the functions $y_\ell(x, \cE)$ are determined systematically by the difference equation \eqref{eq:diff-eq} order by order.
Once these functions are known, Dunham claimed in \cite{Dunham} that the quantum corrections to the BS condition is
incorporated as a formal power series of $\hbar$,
\be
\sum_{\ell=0}^\infty \hbar^\ell \oint_B \rd x \, y_\ell(x, \cE)=2\pi \hbar n.
\label{eq:Dunham-1}
\ee
At the leading order, since the function $y_0(x, \cE)$ satisfies the same algebraic relation as the mirror curve \eqref{eq:mirror-B3},
the left hand side at $\ell=0$ in \eqref{eq:Dunham-1} is just the same one as the LHS in \eqref{eq:BS}.
It turns out that the first correction at $\ell=1$ leads to the constant contribution, $-\pi \hbar$.
This effectively shifts the RHS from $2\pi \hbar n$ to $2\pi \hbar (n+1/2)$.
This is essentially the same story as the WKB approximation in ordinary quantum mechanics.
We conclude that if the sum \eqref{eq:Dunham-1} is truncated at $\ell=1$, then the Bohr--Sommerfeld condition \eqref{eq:BS} is reproduced.

Since the higher odd power corrections $y_{2m+1}(x,\cE)$ turn out to be written as total derivatives of combinations of the even power corrections,
they do not contribute to the periods. 
As a result, the quantization condition \eqref{eq:Dunham-1} is rewritten as
\be
\sum_{m=0}^\infty \hbar^{2m} \oint_B \rd x \, y_{2m}(x,\cE)=2\pi \hbar \( n+\frac{1}{2} \).
\label{eq:Dunham-2}
\ee
It is not easy to perform this integral at each order.
Instead, we use the similar method in the previous subsection.
According to \cite{NS, MiMo}, there exists a deformed version of the prepotential that is implicitly determined by
\be
\ba
t&=\sum_{m=0}^\infty \hbar^{2m} \oint_A \rd x \, y_{2m}(x, \cE), \\
\frac{\pd \cF(t,\hbar)}{\pd t}&=\frac{1}{\hbar} \sum_{m=0}^\infty \hbar^{2m} \oint_B \rd x \, y_{2m}(x,\cE).
\ea
\label{eq:QG}
\ee
We call these the quantum geometry relations because they are actually a quantum deformation of the special geometry relations
\eqref{eq:cB} and \eqref{eq:cA}.
The remarkable fact suggested in \cite{NS} is that this deformed prepotential $\cF(t, \hbar)$ is directly related to the refined topological
string partition function $Z_\text{ref}(t;\epsilon_1, \epsilon_2)$ (or the Nekrasov partition function on the $\Omega$-background
in the gauge theoretic language) in a special limit:
\be
\cF(t ,\hbar)=\lim_{\epsilon_2 \to 0} \epsilon_2 \log Z_\text{ref}(t;\epsilon_1=\ri \hbar, \epsilon_2).
\label{eq:F-NS}
\ee
We refer to this limit as the Nekrasov--Shatashvili (NS) limit and to the deformed prepotential $\cF(t, \hbar)$ as the NS free energy below.
The main advantage of the relation \eqref{eq:F-NS} is that  one can use the powerful technique on the A-model side, 
the so-called topological vertex formalism \cite{AKMV, IKV}.
This method allows us to compute the free energy around the large radius point ($t=\infty$) for any given toric diagram in principle.
Each coefficient in the instanton expansion is a function of the Planck constant $\hbar$ (and the other mass parameters).
This means that the $\hbar$-correction is automatically resummed in all orders at each instanton sector.

For concreteness, let us see our interested case.
As will be reviewed in Appendix~\ref{sec:RTV}, the instanton expansion of the NS free energy for local $\cB_3$ is systematically computed by using
the refined topological vertex.
Up to two instantons, its explicit form is
\be
\ba
&F_\text{NS}^\text{inst}(t,m_i, \hbar)=-\(\frac{1}{m_1}+\frac{1}{m_2}+\frac{1}{m_3}+m_1 m_2+m_2 m_3+m_3 m_1 \)\frac{\re^{-t}}{2\sin \frac{\hbar}{2} }\\
&\quad -\frac{1}{8}\biggl[ 8(m_1+m_2+m_3)\cot \frac{\hbar}{2}  \\ 
&\quad- \biggl( \frac{1}{m_1^2}+\frac{1}{m_2^2}+\frac{1}{m_3^2} 
+m_1^2 m_2^2+m_2^2 m_3^2+m_3^2 m_1^2 \biggr) \frac{1}{\sin \hbar} \biggr]\re^{-2t}+\cO(\re^{-3t}). 
\ea
\label{eq:F-NS-B3}
\ee
Of course, in the classical limit $\hbar \to 0$, this reduces to the prepotential \eqref{eq:F0}: $F_\text{NS}^\text{inst}(t,\hbar) =\hbar^{-1} F_0^\text{inst}(t)+\cO(\hbar)$.
The first quantum correction to $F_\text{NS}^\text{inst}(t,\hbar)$ is also read off from \eqref{eq:F-NS-B3}:
\be
\ba
F_\text{NS}^\text{(1),inst}(t)&=-\frac{1}{24}\(\frac{1}{m_1}+\frac{1}{m_2}+\frac{1}{m_3}+m_1 m_2+m_2 m_3+m_3 m_1 \)\re^{-t}\\
&\quad +\frac{1}{48}\biggl( 8(m_1+m_2+m_3)+\frac{1}{m_1^2}+\frac{1}{m_2^2}+\frac{1}{m_3^2} \\
&\quad +m_1^2 m_2^2+m_2^2 m_3^2+m_3^2 m_1^2 \biggr)\re^{-2t}+\cO(\re^{-3t}). 
\ea
\ee
Using the result in \cite{HKP}, this function can be written in closed form
\be
F_\text{NS}^\text{(1)}(t)=-\frac{1}{24} \log \Delta(z_0).
\label{eq:F-NS-1}
\ee
where $\Delta(z):=g_2^3-27g_3^2$ is the modular discriminant  of the Weierstrass form \eqref{eq:elliptic-curve}.
The parameters $z_0$ and $t$ are related by the usual A-period \eqref{eq:A-period} (not by the quantum one).
To distinguish it from the quantum A-period \eqref{eq:qA-period}, we denote the subscript $0$ in \eqref{eq:F-NS-1}.
Eliminating $z_0$, we indeed obtain the instanton expansion
\be
\ba
F_\text{NS}^\text{(1)}(t)&=-\frac{t}{4}-\frac{\log{(m_1 m_2 m_3)}}{24}\\
&\quad-\frac{1}{24}\(\frac{1}{m_1}+\frac{1}{m_2}+\frac{1}{m_3}+m_1 m_2+m_2 m_3+m_3 m_1 \)\re^{-t}+\cdots.
\ea
\label{eq:F-NS-1-total}
\ee

Let us turn to Dunham's quantization condition.
Using the relation \eqref{eq:QG}, the quantization condition \eqref{eq:Dunham-2} is written as
\be
\frac{\pd \cF(t, \hbar)}{\pd t}=2\pi \( n+\frac{1}{2} \).
\ee
This quantization condition should be understood as a formal power series in $\hbar$,
\be
\frac{\pd}{\pd t}\sum_{m=0}^\infty \hbar^{2m-1} \cF_m(t) = 2\pi \( n+\frac{1}{2} \).
\ee
In general the semiclassical expansion in this equation is a divergent series, and to use it for finite $\hbar$
one needs resummations like the Borel sum.

In the spirit of \cite{NS}, there is an interesting resummation procedure.
In the large radius limit $t \to \infty$, each coefficient $\cF_m(t)$ admits the instanton-like expansion in $\re^{-t}$.
If exchanging the two infinite sums naively, one can first perform the sum for $\hbar$ in each instanton sector.
Then, we obtain the following condition,
\be
\frac{3t^2}{\hbar}-\frac{\log(m_1 m_2 m_3)}{\hbar}t+\frac{C_0}{\hbar}+C_1 \hbar
+\frac{\pd F_\text{NS}^\text{inst}(t,m_i, \hbar)}{\pd t} = 2\pi \( n+\frac{1}{2} \).
\label{eq:QC-naive}
\ee
where $C_0$ is given by \eqref{eq:C0} and $C_1$ is simply given by
\be
C_1=-\frac{1}{4}.
\ee
This constant originates from the first quantum correction \eqref{eq:F-NS-1-total}.
The quantization condition \eqref{eq:QC-naive} is a naive generalization of the original proposal of Nekrasov and Shatashvili in \cite{NS}.
For integrable models associated with 4d supersymmetric gauge theories, such quantization conditions correctly
reproduces the exact eigenvalues (see \cite{KT, HM} for instance).

However, in our case, if we try to use the naive quantization condition \eqref{eq:QC-naive} for finite $\hbar$, we encounter a serious problem.
For instance, the one-instanton term in the NS free energy \eqref{eq:F-NS} has the factor
$1/\sin (\hbar/2)$, and this is obviously singular at $\hbar=2\pi m$ ($m \in \mathbb{Z}$).
The two-instanton term also diverges at $\hbar=\pi m$.
Similarly, the $n$-instanton part is always singular at $\hbar=2\pi m/n$.%
\footnote{This is a reflection of the fact that the instanton expansion of the NS free energy (or of the refined free energy) is well-defined only for $|q_1|<1$
or $|q_1|>1$, where $q_1=\re^{\epsilon_1}=\re^{\ri \hbar}$. It is ill-defined on the unit circle $|q_1|=1$.
The same problem happens in the so-called compact quantum dilogarithm. It is well-known that there is another quantum dilogarithm,
called the non-compact (or Faddeev's) quantum dilogarithm. The compact quantum dilogarithm is not defined for $|q|=1$, 
while the non-compact one is completely well-defined even for $|q|=1$. The structure of the non-compact quantum dilogarithm
is very similar to the left hand side in \eqref{eq:EQC} \cite{Hatsuda-resurgence}.}
It is clear that this problem is caused by the exchange of the sums for $\hbar$ and for $\re^{-t}$. 
This problem was pointed out in \cite{KaMa} in the context of ABJM theory, based on the earlier works \cite{HMO2, HMMO}.
The idea in \cite{KaMa} to resolve this problem is that there are additional contributions that
look like quantum mechanically nonperturbative corrections in the large $t$ instanton expansion.

Based on \cite{HW, GHM1}, it was conjectured in \cite{WZH} that the correct eigenvalues are determined by the following quantization condition
\be
\frac{3t^2}{\hbar}-\frac{\log(m_1 m_2 m_3)}{\hbar}t+\frac{C_0}{\hbar}+C_1 \hbar
+\frac{\pd F_\text{NS}^\text{inst}(t,m_i, \hbar)}{\pd t}+\frac{\pd F_\text{NS}^\text{inst}(\wt{t},\wt{m}_i, \wt{\hbar})}{\pd \wt{t}}=2\pi \( n+\frac{1}{2} \),
\label{eq:EQC}
\ee
where
\be
\wt{t}=\frac{2\pi t}{\hbar}, \qquad \wt{m}_i=m_i^{2\pi/\hbar}, \qquad \wt{\hbar}=\frac{4\pi^2}{\hbar}.
\ee
This is the main result in this review part.
Since the quantization condition \eqref{eq:EQC} includes all the quantum corrections in $\hbar$,
it is regarded as an exact version of the Bohr--Sommerfeld quantization condition \eqref{eq:BS-2}.
We refer to \eqref{eq:EQC} as the \textit{exact quantization condition}.
The singular part of $\pd F_\text{NS}^\text{inst}(t,m_i, \hbar)/\pd t$ at $\hbar=2\pi m/n$ is always canceled by
the singular part of the ``dual'' contribution $\pd F_\text{NS}^\text{inst}(\wt{t},\wt{m}_i, \wt{\hbar})/\pd \wt{t}$ at $\wt{\hbar}=2\pi n/m$.
The left hand side of \eqref{eq:EQC} is totally well-defined for any $\hbar$.
Since the dual part is the expansion of $\re^{-\wt{t}}=\re^{-2\pi t/\hbar}$, it looks nonperturbative in $\hbar$.
We emphasize that such nonperturbative corrections do not appear in the 4d case.
They are needed only for eigenvalue problems associated with 5d gauge theories/topological strings.

As in the previous subsection, the quantization condition \eqref{eq:EQC} fixes the K\"ahler modulus $t$ for a given non-negative integer $n$.
To know the eigenvalue $\cE$, we need the quantum version of the A-period given by the first equation in \eqref{eq:QG}.
As reviewed in Appendix~\ref{sec:RTV}, one can compute the large $\cE$ expansion of this quantum A-period from the difference equation \eqref{eq:diff-eq}.
In the current case, the quantum A-period admits the following small $z$ expansion:
\be
\ba
-t&=\log z+\Pi_A(z, m_i, \hbar),\\
\Pi_A(z,m_i, \hbar)&=(m_1+m_2+m_3)z^2+(q^{1/2}+q^{-1/2})(1+m_1 m_2 m_3)z^3 \\
&\hspace{-1.5cm} +\biggl[ \frac{3}{2}(m_1^2+m_2^2+m_3^2)+(4+q+q^{-1})(m_1 m_2+m_2 m_3+m_3 m_1) \biggr] z^4+\cO(z^5),
\ea
\label{eq:qA-period}
\ee
or inversely
\be
\ba
\cE^{-1}&=z=Q-(m_1+m_2+m_3)Q^3-(q^{1/2}+q^{-1/2})(1+m_1 m_2 m_3)Q^4\\
&+[m_1^2+m_2^2+m_3^2-(q+q^{-1}-1)(m_1 m_2 +m_2 m_3+m_3 m_1)]Q^5+\cO(Q^6).
\ea
\label{eq:q-mirror-map}
\ee
Using this relation, we can recover the original energy $\cE$.

Finally, we compute the semiclassical expansions of the eigenvalues.
Originally, Dunham's quantization condition is given by a formal power series in $\hbar$, as in \eqref{eq:Dunham-2}.
It is convenient to use this form to derive the semiclassical expansion of $\cE$.
Let us denote the quantum correction to the B-period by
\be
\vol_m(\cE) := \oint_B \rd x \, y_{2m}(x, \cE).
\ee
Then Dunham's condition is written as
\be
\sum_{m=0}^\infty \hbar^{2m} \vol_{m}(\cE_n)=2\pi \hbar \( n+\frac{1}{2} \)
\label{eq:WKB-QC}
\ee
We want to know the small $\hbar$ expansion of $\cE_n$:
\be
\cE_n=\cE_n^{(0)}+\hbar \cE_n^{(1)}+\hbar^2 \cE_n^{(2)}+\hbar^3 \cE_n^{(3)}+\cdots.
\ee
Plugging this expansion into the quantization condition \eqref{eq:WKB-QC},
we find the condition at the leading order:
\be
\vol_0(\cE_n^{(0)})=0.
\ee
This happens if and only if $\cE_n^{(0)}=6$ (for $m_1=m_2=m_3=1$).
In fact, the region surrounded by the mirror curve \eqref{eq:mirror-B3} shrinks to a point for $\cE=6$.
In other words, the mirror curve $\cE=\re^{x}+\re^{-x}+\re^{y}+\re^{-y}+\re^{x+y}+\re^{-x-y}$
takes the minimal value $\cE=6$ at $x=y=0$.
At the next-to-leading order, we also obtain the relation
\be
\cE_n^{(1)} \vol_0'(6)=2\pi \(n+\frac{1}{2} \).
\ee
Therefore to get the first correction $\cE_n^{(1)}$, we need $\vol_0'(6)$.
This derivative is directly evaluated by the formula \eqref{eq:classical-periods-2}.
Since we have
\be
\vol_0'(\cE)=\frac{8}{\sqrt{\cE^2-12+8\sqrt{\cE+3}}}
\eK \( \frac{\cE^2-12-8\sqrt{\cE+3}}{\cE^2-12+8\sqrt{\cE+3}} \),
\label{eq:vol0'}
\ee
we find
\be
\vol_0'(6)=\frac{\pi}{\sqrt{3}} \quad \Rightarrow \quad \cE_n^{(1)}=\sqrt{3}(2n+1).
\ee
To compute $\cE_n^{(2)}$ and $\cE_n^{(3)}$, we need $\vol_1(\cE)$.
Interestingly, the quantum correction $\vol_m(\cE)$ is generated by acting a differential operator on the classical period:
\be
\vol_m(\cE)=\cD_m \vol_0(\cE).
\ee
As we will see in appendix~\ref{sec:E8}, the correspondence to the mass deformed $E_8$ del Pezzo geometry helps us to compute
the first differential operator (see \cite{HKRS}),
\be
\cD_1=\frac{2-\cE}{24}\pd_\cE+\frac{12+2\cE-\cE^2}{24} \pd_\cE^2.
\ee 
From \eqref{eq:vol0'}, $\vol_1(\cE)$ is thus written in closed form.
Using these results, we finally obtain the following semiclassical expansion of $\cE_n$ up to $\hbar^3$:
\be
\cE_n=6+\sqrt{3}(2n+1)\hbar+\frac{2n^2+2n+1}{4}\hbar^2+\frac{4n^3+6n^2+8n+3}{72\sqrt{3}} \hbar^3+\cO(\hbar^4).
\label{eq:E-WKB}
\ee
This implies that all the eigenvalues degenerate to $\cE=6$ in the classical limit $\hbar \to 0$.

\subsection{Testing the quantization condition}\label{subsec:test}
In the previous subsection, we wrote down the exact quantization condition \eqref{eq:EQC} that solves the eigenvalue problem \eqref{eq:diff-eq}.
Since this quantization condition is not derived from first principles (or not proved rigorously), one has to confirm its validity.
The most direct way is to compare the eigenvalues obtained by \eqref{eq:EQC} with the numerical ones obtained
by diagonalizing the Hamiltonian \eqref{eq:H-B3}.
In the following, for simplicity, we check it for the most symmetric case:
\be
m_1=m_2=m_3=1.
\ee
The other values of $(m_1,m_2, m_3)$ can be checked in the similar way.

We first consider the case $\hbar=2\pi$.
This is the special case that the dual Planck constant $\wt{\hbar}$ equals to $\hbar$.
Using the explicit form of the NS free energy \eqref{eq:F-NS-B3}, one finds that the quantization condition
is drastically simplified. One can see
\be
\ba
&\lim_{\hbar \to 2\pi}\( \frac{\pd F_\text{NS}^\text{inst}(t,m_i=1, \hbar)}{\pd t}+\frac{\pd F_\text{NS}^\text{inst}(\wt{t},\wt{m}_i=1, \wt{\hbar})}{\pd \wt{t}}\) \\
&=\frac{1}{2\pi} \left[ 6(t+1)\re^{-t}-\frac{21}{2}(2t+1)\re^{-2t}+\frac{56}{3}(3t+1)\re^{-3t}+\cO(t \re^{-4t}) \right].
\ea
\ee
This is written in terms of the prepotential $F_0^\text{inst}(t)$ as
\be
\frac{1}{2\pi} [t \pd_t^2 F_0^\text{inst}(t+\pi \ri)-\pd_t F_0^\text{inst}(t+\pi \ri)].
\ee
We conclude that the exact quantization condition at $\hbar=2\pi$ is finally given by
\be
3t^2-2\pi^2+t \pd_t^2 F_0^\text{inst}(t+\pi \ri)-\pd_t F_0^\text{inst}(t+\pi \ri)=4\pi^2 \( n+\frac{1}{2} \).
\label{eq:EQC-2pi}
\ee
The same equation for local $\mathbb{P}^2$ was originally found in \cite{GHM1},
and this form is almost universal for other CYs.
Therefore for $\hbar=2\pi$, the only building block of the quantization condition is the prepotential $F_0^\text{inst}(t)$,
which is computed up to any order by using the formula \eqref{eq:classical-periods}.
The relation between $t$ and $\cE$ is given by the quantum mirror map \eqref{eq:qA-period} or \eqref{eq:q-mirror-map}.
For $\hbar=2\pi$, it gives
\be
z=Q-3Q^3+4Q^4-32Q^7+144Q^8+\cO(Q^9).
\label{eq:q-mirror-map-2pi}
\ee
This is formally equivalent to the classical mirror map \eqref{eq:mirror-map} by replacing $z \to -z$ and $Q \to -Q$.
This replacement is a reflection that the argument of the prepotential in \eqref{eq:EQC-2pi} is shifted by $t \to t +\pi \ri$.

Now we compare the eigenvalues computed by \eqref{eq:EQC-2pi} and \eqref{eq:q-mirror-map-2pi} with
the numerical values.
A numerical method to compute the eigenvalues of $H$ is explained in \cite{HW}.
The idea is very simple. We represent $H$ in the harmonic oscillator basis.
The matrix elements $\langle n |H| m \rangle$ is easily computed.
Here $\ket{n}$ is the $n$-th Fock state.
This matrix is infinite dimensional, but we truncate it as an $(L+1) \times (L+1)$ matrix.
If one takes sufficiently large $L$, the eigenvalues of the truncated matrix should give
a good approximation of the true eigenvalues.
In this way, one can evaluate the numerical values of the eigenvalues with high precision.%
\footnote{In the practical calculation, it is much more convenient to use the canonical transformed
Hamiltonian \eqref{eq:H-B3-shift} because in this representation the matrix elements are non-vanishing
only for $n \equiv m \pmod{6}$. This fact drastically improves the speed of convergence in $L \to \infty$.}

In Tables~\ref{tab:E0-B3-2pi} and \ref{tab:E1-B3-2pi}, we show the eigenvalues at $\hbar=2\pi$ for the ground state
and the first excited state, respectively.
In solving the quantization condition \eqref{eq:EQC-2pi}, we truncate the instanton sum up to some order of $Q=\re^{-t}$.
We show the values for several maximal orders.
It is obvious to see that the higher instanton corrections improve the accuracy of the eigenvalues. 
It is observed that the ground state eigenvalue here equals to that for the mass deformed $E_8$ del Pezzo geometry
with particular mass parameters $(M_1, M_2, M_3)=(2, 3, 3)$
even though these two eigenvalue problems look quite different. See Table 4.14 in \cite{GKMR} and appendix~\ref{sec:E8}. 

\begin{table}[tb]
\caption{The ground state eigenvalue for $\hbar=2\pi$.}
\begin{center}
  \begin{tabular}{cl}
\hline
Maximal order &  \hspace{2cm}$E_0=\log \cE_0$ \\
\hline
$Q^5$   & ${\bf 3.597651}38849044946015022964047$\\
$Q^{10}$ & ${\bf 3.5976516128}1389105626031597607$\\
$Q^{15}$ & ${\bf 3.597651612809098}45859613439611$\\
$Q^{20}$ & ${\bf 3.59765161280909860323}830723863$\\
$Q^{25}$ & ${\bf 3.597651612809098603233253}01290$\\
$Q^{30}$ & ${\bf 3.5976516128090986032332532044}3$\\
\hline
Numerical value & ${\bf 3.59765161280909860323325320442}$ \\
\hline
  \end{tabular}
\end{center}
\label{tab:E0-B3-2pi}
\end{table}
\begin{table}[tb]
\caption{The first excited eigenvalue for $\hbar=2\pi$.}
\begin{center}
  \begin{tabular}{cl}
\hline
Maximal order &  \hspace{2.5cm}$E_1=\log \cE_1$ \\
\hline
$Q^{5}$   & ${\bf 5.1233244150}319203600284921267610931$\\
$Q^{10}$ & ${\bf 5.123324415056734318}5793231592826966$\\
$Q^{15}$ & ${\bf 5.123324415056734318320823}5981093662$\\
$Q^{20}$ & ${\bf 5.123324415056734318320823601884633}2$\\
\hline
Numerical value & ${\bf 5.1233244150567343183208236018846331}$ \\
\hline
  \end{tabular}
\end{center}
\label{tab:E1-B3-2pi}
\end{table}

For other values of $\hbar$, we need to solve the quantization condition \eqref{eq:EQC}.
To do so, we computed $F_\text{NS}^\text{inst}(t,\hbar)$ up to $Q^{15}$ by using the refined topological vertex.
We also computed the quantum mirror map \eqref{eq:q-mirror-map} up to the same order.
Using these data, we solve the condition \eqref{eq:EQC} for $\hbar \ne 2\pi$.
In Tables~\ref{tab:E0-B3-6}, we show the eigenvalues for $\hbar=6$.
As in the case of $\hbar=2\pi$, our quantization condition reproduces the correct answer of the problem.
For $\hbar=6$, one can also solve the naive guess of the quantization condition \eqref{eq:QC-naive}.
If using the instanton expansion up to $Q^{12}$, we obtain the following value of the logarithm of the ground state energy
\be
E_0^\text{naive}(\hbar=6)=\log \cE_0^\text{naive}=3.343\cdots.
\ee
This is far from the correct eigenvalue $E_0(\hbar=6)=3.51693\cdots$.
All of these tests strongly support the validity of our quantization condition \eqref{eq:EQC}.

\begin{table}[tb]
\caption{The first two eigenvalues for $\hbar=6$. We solved the quantization condition \eqref{eq:EQC},
and used the quantum mirror map \eqref{eq:q-mirror-map}.
The maximal orders $(Q^n,\wt{Q}^m)$ mean that we use the instanton expansions of $F_\text{NS}^\text{inst}(t, \hbar)$
up to order $Q^n$ and of $F_\text{NS}^\text{inst}(\wt{t}, \wt{\hbar})$ up to order $\wt{Q}^m$.}
\begin{center}
  \begin{tabular}{cll}
\hline
Maximal orders &  \qquad \; $E_0=\log \cE_0$ & \qquad \qquad\quad $E_1=\log \cE_1$ \\
\hline
$(Q^{3}, \wt{Q}^{3})$     & ${\bf 3.5169}1182900199066$ & ${\bf 5.0072093}39313409829211209289$\\
$(Q^{6}, \wt{Q}^{6})$     & ${\bf 3.51693}300839154630$ & ${\bf 5.00720939007}5040706862230424$\\
$(Q^{9}, \wt{Q}^{9})$     & ${\bf 3.5169329855}3663716$ & ${\bf 5.0072093900744463}84927798130$\\
$(Q^{12}, \wt{Q}^{12})$ & ${\bf 3.5169329855670}4748$ & ${\bf 5.007209390074446393505}480957$\\
$(Q^{15}, \wt{Q}^{15})$ & ${\bf 3.51693298556700}599$ & ${\bf 5.00720939007444639350535266}3$\\
\hline
Numerical values & ${\bf 3.51693298556700604}$ & ${\bf 5.007209390074446393505352665}$\\
\hline
  \end{tabular}
\end{center}
\label{tab:E0-B3-6}
\end{table}

\section{Local $\cB_3$ geometry and electrons on 2d lattice}\label{sec:CY-electrons}
In the previous section, we reviewed the eigenvalue problem appearing in
the quantum geometry of local $\cB_3$.
Our conclusion is that the eigenvalue problem is solved by the exact quantization condition \eqref{eq:EQC}.
Here we will explain how this problem meets a two-dimensional electron system with a magnetic flux.

\subsection{Strong-weak relations}
The quantization condition \eqref{eq:EQC} has a remarkable symmetric property.
It is invariant under 
\be
(t, m_i, \hbar) \mapsto (\wt{t},\wt{m}_i, \wt{\hbar})=\( \frac{2\pi t}{\hbar}, m_i^{2\pi/\hbar}, \frac{4\pi^2}{\hbar} \).
\label{eq:S-transform}
\ee
In fact, it is obvious to see that the instanton correction $\pd F_\text{NS}^\text{inst}(t,m_i, \hbar)/\pd t$ and
its dual $\pd F_\text{NS}^\text{inst}(\wt{t},\wt{m}_i, \wt{\hbar})/\pd \wt{t}$ are exchanged by this transformation.
Also one can show the following equality
\be
\frac{3\wt{t}^2}{\wt{\hbar}}-\frac{\log(\wt{m}_1 \wt{m}_2 \wt{m}_3)}{\wt{\hbar}}\wt{t}+\frac{\wt{C}_0}{\wt{\hbar}}+\wt{C}_1 \wt{\hbar}
=\frac{3t^2}{\hbar}-\frac{\log(m_1 m_2 m_3)}{\hbar}t+\frac{C_0}{\hbar}+C_1 \hbar
\ee
where
\be
\wt{C}_0=-\pi^2-\frac{\log^2 \wt{m}_1+\log^2 \wt{m}_2+\log^2 \wt{m}_3}{2},\qquad
\wt{C}_1=-\frac{1}{4}.
\ee
This invariance implies that the system at $\hbar$ are related to the system at $\wt{\hbar}$.
Since these two couplings satisfy
\be
\hbar \wt{\hbar} =4\pi^2,
\ee
the strong coupling regime of $\hbar$-system is mapped to the weak coupling regime of $\wt{\hbar}$-system and vice versa.
In this sense, the quantum mechanical system \eqref{eq:diff-eq} has a self S-dual structure with respect to the Planck constant.%
\footnote{Though the Calabi--Yau manifold itself does not change in the transformation \eqref{eq:S-transform}, 
the mass parameter $\wt{m}_i$ in the dual system differs from the original one $m_i$ except for $m_i=1$.}
This property was particularly emphasized in \cite{Hatsuda-EQC}.
In the relativistic Toda lattice \cite{KLS, Sciarappa}, such an S-duality is related to the so-called modular double duality
in quantum group \cite{Faddeev1, Faddeev2}.
The special case of $\hbar=\wt{\hbar}=2\pi$ is just the fixed point.
Recalling that the K\"ahler modulus is a function of $\cE$, $m_i$ and $\hbar$, as in \eqref{eq:qA-period},
the above transformation rule relates the energy $\wt{\cE}$ of the $\wt{\hbar}$-system 
to $\cE$ of the $\hbar$-system.
More explicitly, the equation
\be
\wt{t}(\wt{\cE},\wt{m}_j, \wt{\hbar} )=\frac{2\pi}{\hbar} t(\cE, m_j, \hbar),
\label{eq:t-Sdual}
\ee
relates $\wt{\cE}$ to $\cE$.

Let us see a consequence of the relation \eqref{eq:t-Sdual}.
In the following, we again restrict our attention to the case of $m_1=m_2=m_3=1$.
Let us consider the case of $\hbar=\pi$ and $\wt{\hbar}=4\pi$.
From the quantum A-period \eqref{eq:qA-period}, we have
\be
\ba
-t(\cE, \hbar=\pi)&=\log z+3z^2+\frac{21}{2}z^4+56z^6+\frac{1485}{4} z^8+\frac{14058}{5}z^{10}+\cO(z^{12}), \\
-\wt{t}(\wt{\cE}, \wt{\hbar}=4\pi)&=\log \wt{z}+3\wt{z}^2+4\wt{z}^3+\frac{45}{2}\wt{z}^4+72\wt{z}^5
+340 \wt{z}^6+1440 \wt{z}^7\\
&\quad+\frac{27405}{4}\wt{z}^8+\frac{96880}{3}\wt{z}^9++\frac{794178}{5}\wt{z}^{10}+\cO(\wt{z}^{11}),
\ea
\ee
where
\be
z=\frac{1}{\cE},\qquad \wt{z}=\frac{1}{\wt{\cE}}.
\ee
One can see that if and only if the energies satisfy the relation
\be
\wt{\cE}=\cE^2-6,
\label{eq:E-pi-4pi}
\ee
then the two moduli satisfy $\wt{t}(\wt{\cE},4\pi)=2t(\cE,\pi)$.
Conversely speaking, the requirement $\wt{t}(\wt{\cE},4\pi)=2t(\cE,\pi)$ leads to the energy relation \eqref{eq:E-pi-4pi}.
In this way, one can find explicit relations between $\cE$ and $\wt{\cE}$.
In Table~\ref{tab:rel-B3}, we show these relations for various $\hbar$ of the form $2\pi a/b$,
where $a$ and $b$ are coprime integers.

\begin{table}[tb]
\caption{The strong-weak energy relation in local $\cB_3$ for $m_1=m_2=m_3=1$. 
The energy $\cE$ at $\hbar=2\pi a/b$ is related to the energy $\wt{\cE}$ at $\wt{\hbar}=2\pi b/a$
by the algebraic equation $F_{b/a}(\wt{\cE})=F_{a/b}(\cE)$.
We found these relations from the equation \eqref{eq:t-Sdual}.}
\begin{center}
  \begin{tabular}{ccl}
\hline
$a$ & $b$ &  $F_{b/a}(\wt{\cE})=F_{a/b}(\cE)$ \\
\hline
$1$ & $1$ & $\wt{\cE}=\cE$ \\
      & $2$ & $\wt{\cE}=\cE^2-6$ \\
      & $3$ & $\wt{\cE}=\cE^3-9\cE-6$ \\
      & $4$ & $\wt{\cE}=\cE^4-12\cE^2-8\sqrt{2}\cE+6$ \\
      & $5$ & $\wt{\cE}=\cE^5-15\cE^3-5(1+\sqrt{5})\cE^2+\frac{15}{2}(5-\sqrt{5})\cE+15(-1+\sqrt{5})$ \\
      & $6$ & $\wt{\cE}=\cE^6-18\cE^4-12\sqrt{3}\cE^3+45\cE^2+36\sqrt{3}\cE+6$ \\ \hline
$2$ & $3$ & $\wt{\cE}^2-6=\cE^3-9\cE+6$ \\
      & $5$ & $\wt{\cE}^2-6=\cE^5-15\cE^3+5(1-\sqrt{5})\cE^2+\frac{15}{2}(5+\sqrt{5})\cE+15(1+\sqrt{5})$ \\ \hline
$3$ & $4$ & $\wt{\cE}^3-9\wt{\cE}+6=\cE^4-12\cE^2+8\sqrt{2}\cE+6$ \\
      & $5$ & $\wt{\cE}^3-9\wt{\cE}-6=\cE^5-15\cE^3+5(-1+\sqrt{5})\cE^2+\frac{15}{2}(5+\sqrt{5})\cE-15(1+\sqrt{5})$ \\ \hline
$4$ & $5$ & $\wt{\cE}^4-12\wt{\cE}^2+8\sqrt{2}\wt{\cE}+6$ \\
       &       & $=\cE^5-15\cE^3+5(1+\sqrt{5})\cE^2+\frac{15}{2}(5-\sqrt{5})\cE+15(1-\sqrt{5})$ \\ \hline 
$5$ & $6$ & $\wt{\cE}^5-15\wt{\cE}^3+5(1+\sqrt{5})\wt{\cE}^2+\frac{15}{2}(5-\sqrt{5})\wt{\cE}+15(1-\sqrt{5})$ \\
       &       & $=\cE^6-18\cE^4+12\sqrt{3}\cE^3+45\cE^2-36\sqrt{3}\cE+6$ \\
\hline
  \end{tabular}
\end{center}
\label{tab:rel-B3}
\end{table}

In general, the relation \eqref{eq:t-Sdual} for $\hbar=2\pi a/b$ leads to an algebraic equation
\be
F_{b/a}(\wt{\cE})=F_{a/b}(\cE),
\label{eq:strong-weak}
\ee
where $F_{a/b}(\cE)$ is a polynomial with degree $b$.
It is not easy to find the general form of $F_{a/b}(\cE)$ from \eqref{eq:t-Sdual} only.
However, using the connection with a 2d electron system in the next subsection,
we can conjecture $F_{a/b}(\cE)$. 
Here we write only the final result:
\be
F_{a/b}(\cE)=D_{a/b}(\cE)+2(1+(-1)^b+(-1)^{(a-1)b} ),
\label{eq:Fab}
\ee
where
\be
D_{a/b}(\cE)=\det \begin{pmatrix}
\cA_1 & \cB_1   & 0   & \cdots & 0 & 0 & \cB_b^* \\
\cB_1^*  &  \cA_2 & \cB_2 & \cdots & 0 & 0 & 0 \\
\vdots  & \vdots & \vdots  &  & \vdots & \vdots & \vdots \\
0  &  0  &  0  &  \cdots  &  \cB_{b-2}^*  &  \cA_{b-1}  &  \cB_{b-1}  \\
\cB_b   &  0  &  0  &  \cdots  &  0  &  \cB_{b-1}^*  &  \cA_b  
\end{pmatrix},
\ee
and
\be
\cA_j := \cE+2\cos (2\pi \tau j),\qquad \cB_j := -(1-\re^{-2\pi \ri \tau j} ),
\qquad \tau=\frac{a}{b}.
\ee
This relies on the known result in condensed matter physics \cite{HK}.
We have confirmed that the formula \eqref{eq:Fab} reproduces all of the relations in Table~\ref{tab:rel-B3}
except for $b=2$. It seems that for $b=2$ this formula does not work, but in this case, we easily find
\be
F_{a/2}(\cE)=\cE^2-6 \qquad (\text{odd }a).
\ee

As in \cite{HKT}, the same relation is obtained from the difference equation \eqref{eq:diff-eq} and its dual.
To see this, it is more convenient to shift the variable $\y \to \y-\x/2$.
Then the Hamiltonian is 
\be
H=\re^{\x}+\re^{-\x}+\re^{-\frac{\x}{2}+\y}+\re^{\frac{\x}{2}-\y}+\re^{-\frac{\x}{2}-\y}+\re^{\frac{\x}{2}+\y}
\label{eq:H-B3-shift}
\ee
Now the difference equation is written as
\be
2\cosh \( \frac{x}{2}+\frac{\ri \hbar}{4} \) \Psi(x+\ri \hbar)
+2\cosh \( \frac{x}{2}-\frac{\ri \hbar}{4} \) \Psi(x-\ri \hbar)
=(\cE-2\cosh x)\Psi(x).
\label{eq:diff-eq-equal-mass}
\ee
We observe that the relations in Table~\ref{tab:rel-B3} are obtained from the compatibility condition between this difference equation and 
its dual:
\be
2\cosh \biggl( \frac{\wt{x}}{2}+\frac{\ri \wt{\hbar}}{4} \biggr) \Psi(x+2\pi \ri)
+2\cosh \biggl( \frac{\wt{x}}{2}-\frac{\ri \wt{\hbar}}{4} \biggr) \Psi(x-2\pi \ri)
=(\wt{\cE}-2\cosh \wt{x})\Psi(x),
\label{eq:diff-eq-equal-mass-dual}
\ee
where
\be
\wt{\hbar}=\frac{4\pi^2}{\hbar}, \qquad \wt{x}=\frac{2\pi x}{\hbar}.
\ee
In the relativistic Toda lattice, it is known that the existence of the dual difference equation like \eqref{eq:diff-eq-equal-mass-dual} 
is a consequence of the modular double duality in the underlying $\cU_q(sl_2(\mathbb{R}))$ symmetry \cite{KLS}.
In our case, the same structure exists.
In fact, the difference equation \eqref{eq:diff-eq-equal-mass-dual} is a consequence of
the dual eigenvalue problem $\wt{H}\Psi(x)=\wt{\cE}\Psi(x)$, where the dual Hamiltonian is defined by
\be
\wt{H}=\re^{\wt{\x}}+\re^{-\wt{\x}}+\re^{-\frac{\wt{\x}}{2}+\wt{\y}}+\re^{\frac{\wt{\x}}{2}-\wt{\y}}
+\re^{-\frac{\wt{\x}}{2}-\wt{\y}}+\re^{\frac{\wt{\x}}{2}+\wt{\y}},
\qquad [\wt{\x}, \wt{\y} ]=\ri \wt{\hbar}.
\ee
The two Hamiltonians commute:
\be
[H, \wt{H}]=0.
\ee
Therefore one can diagonalize these two operators simultaneously,
and it requires the compatibility condition of \eqref{eq:diff-eq-equal-mass} and \eqref{eq:diff-eq-equal-mass-dual} for $\hbar=2\pi a/b$. 
It would be interesting to understand the general relation between the quantum geometry for toric CY threefolds
and the modular double duality in quantum group.

\subsection{Analyticity of K\"ahler modulus}
In quantum geometry, the K\"ahler modulus $t(\cE,m_i, \hbar)$ receives quantum corrections as in \eqref{eq:QG},
and it turns out to have a complicated analytic property as a function of $\cE$.
In this subsection, we investigate it, and see that it meets the band spectrum in a 2d electron system
on a triangular lattice with a magnetic flux.

We first observe that the quantum K\"ahler modulus $t(\cE,m_i, \hbar)$ has the following symmetries:
\be
t(\cE,m_i,\hbar+4\pi)=t(\cE,m_i,\hbar)=t(\cE,m_i,-\hbar)=t(\cE,m_i,4\pi-\hbar).
\label{eq:symmetry-t}
\ee
This can be seen by the large radius expansion of the quantum A-period \eqref{eq:qA-period}.
Let us denote the instanton part of the classical A-period by $\Pi_A^{(0)}(\cE)$, i.e.,
\be
\Pi_A^{(0)}(\cE)=\frac{3}{\cE^2}+\frac{4}{\cE^3}+\frac{45}{2\cE^4}+\frac{72}{\cE^5}+\frac{340}{\cE^6}+\cO(\cE^{-7})
\ee
From the periodicity, the K\"ahler modulus for $\hbar=4\pi$ trivially equals to the classical A-period:
\be
t(\cE,4\pi)=\log \cE-\Pi_A^{(0)}(\cE).
\ee
Using \eqref{eq:classical-periods-2}, its derivative is explicitly written as
\be
\frac{\pd t(\cE,4\pi)}{\pd \cE}=\frac{2}{\pi \sqrt{\cE^2-12+8\sqrt{3+\cE}}}
\eK \( \frac{16\sqrt{3+\cE}}{\cE^2-12+8\sqrt{3+\cE}} \).
\ee
We have already seen in the previous section that the quantum A-period for $\hbar=2\pi$ is given by
\be
t(\cE, 2\pi)=\log \cE-\Pi_A^{(0)}(-\cE).
\ee
and
\be
\frac{\pd t(\cE,2\pi)}{\pd \cE}=\frac{2}{\pi \sqrt{\cE^2-12+8\sqrt{3-\cE}}}
\eK \( \frac{16\sqrt{3-\cE}}{\cE^2-12+8\sqrt{3-\cE}} \).
\ee
The important point is that the classical A-period \eqref{eq:classical-periods-2} has branch cuts along
\be
-3 \leq \cE \leq 6.
\ee
Note that $\cE=-3$ and $\cE=6$ are zeros of the discriminant $\Delta=(\cE+2)^3(\cE+3)^2(\cE-6)$ 
for the elliptic curve \eqref{eq:elliptic-curve}.%
\footnote{More precisely, the brunch cut along $-3 \leq \cE < -2$ comes from the square root factor,
while the cut along $-2 < \cE \leq 6$ comes from the complete elliptic integral.
At $\cE=-2$, the derivative of the A-period is divergent.}

Let us consider the analytic property of $t$ at $\hbar=2\pi a/b$.
If $ab$ is even, $\tau=a/b$ is reduced to $\tau=2$, i.e. $\hbar=4\pi$, by combination of the S-transform $\tau \to 1/\tau$,
the T-transform $\tau \to \tau+2$ and the reflection: $\tau \to -\tau$.
By the same argument in \cite{HKT}, we finally obtain
\be
t(\cE,\hbar=2\pi a/b)=\frac{1}{b} \( \log \wt{\cE}-\Pi_A^{(0)}(\wt{\cE}) \) \quad \text{for} \quad \text{$ab$ : even},
\ee
where
\be
\wt{\cE}=F_{a/b}(\cE).
\ee
In this case, the branch cut of $t$ is located along the interval $-3 \leq \wt{\cE} \leq 6$ in the complex $\wt{\cE}$-plane.
This is translated into the condition for $\cE$,
\be
-3 \leq F_{a/b}(\cE)  \leq 6 \quad \text{for} \quad \text{$ab$ : even}.
\ee
If $ab$ is odd, $\tau=a/b$ is reduced to $\tau=1$, i.e. $\hbar=2\pi$. Then the flat coodinate is given by
\be
t(\cE,\hbar=2\pi a/b)=\frac{1}{b} \( \log \wt{\cE}-\Pi_A^{(0)}(-\wt{\cE}) \) \quad \text{for} \quad \text{$ab$ : odd},
\ee
The energy condition for branch cuts is
\be
-6\leq F_{a/b}(\cE)  \leq 3 \quad \text{for} \quad \text{$ab$ : odd}.
\ee

Let us see some concrete examples.
For $(a,b)=(1,1)$ ($\hbar=2\pi$), we have
\be
F_{1/1}(\cE)=\cE.
\ee
Since $ab=1$ is odd, the energy condition is
\be
-6 \leq \cE \leq 3.
\ee

For $(a,b)=(1,2)$ ($\hbar=\pi$), we have
\be
F_{1/2}(\cE)=\cE^2-6.
\ee
Since $ab=2$ is even, the energy condition is
\be
-3 \leq \cE^2-6 \leq 6.
\ee
This is equivalent to
\be
-2\sqrt{3}\leq \cE \leq -\sqrt{3},\qquad \sqrt{3}\leq \cE \leq 2\sqrt{3}.
\ee
In the complex $\cE$-plane, $t(\cE, \pi)$ has two branch cuts along these intervals.

\begin{figure}[t]
\begin{center}
    \includegraphics[width=0.6\linewidth]{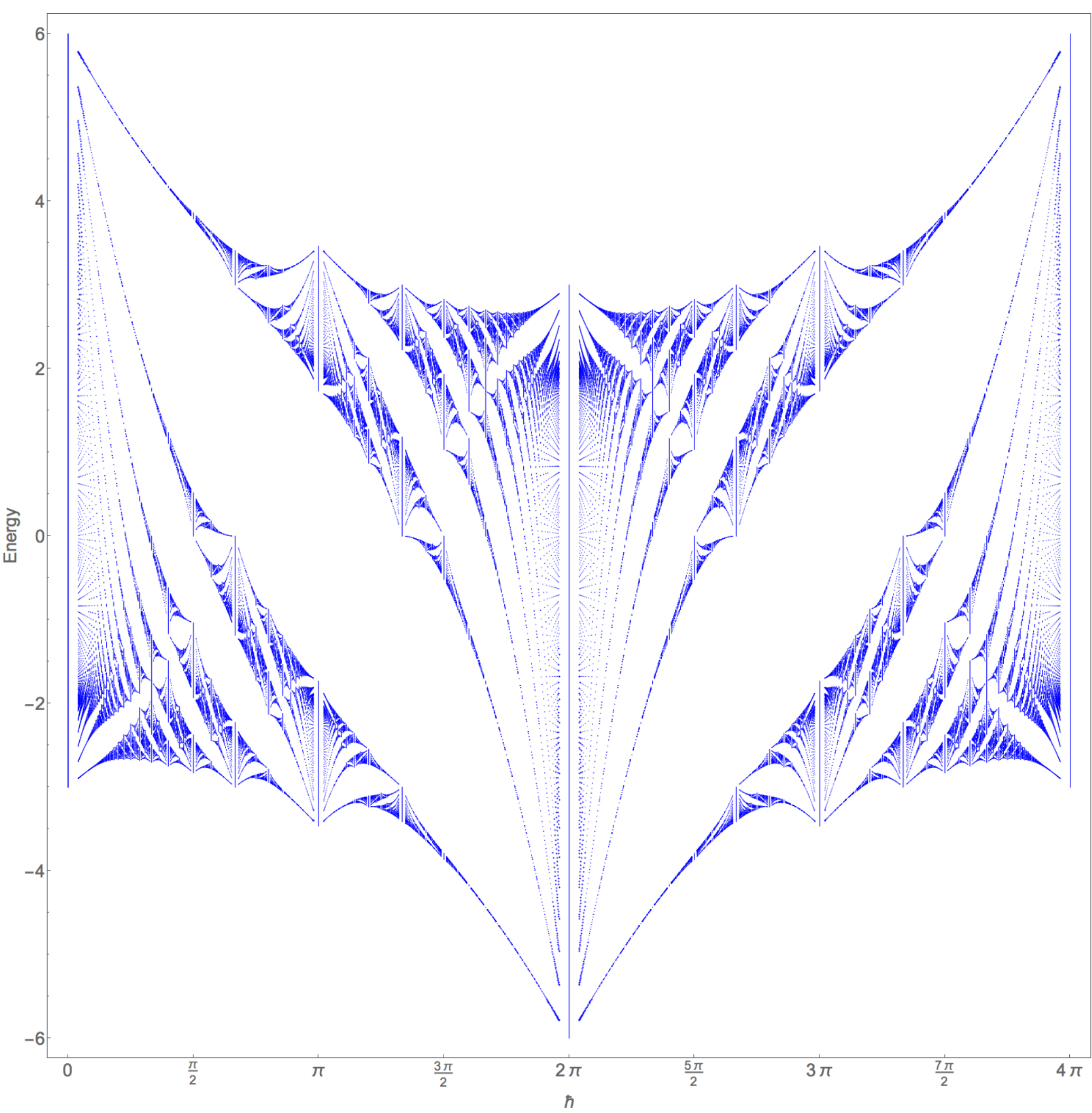} 
\end{center}
  \caption{The branch cut structure of the K\"ahler modulus $t(\cE,\hbar)$ in the quantum local $\cB_3$ geometry with the symmetric masses $m_1=m_2=m_3=1$. The horizontal direction is the Planck constant $\hbar$, and the vertical direction is the energy $\cE$. 
The figure is identical to the band spectrum of electrons on the 2d triangular lattice in the presence of the magnetic field \cite{CW}.
In this context, the horizontal direction corresponds to the magnetic flux $\phi$.}
  \label{fig:cuts-B3-1}
\end{figure}

For $(a,b)=(1,3)$ ($\hbar=\frac{2\pi}{3}$), we have
\be
F_{1/3}(\cE)=\cE^3-9\cE-6,
\ee
and thus the energy condition is
\be
-3\leq \cE\leq -2.22668, \qquad -1.18479\leq \cE\leq 0,\qquad 3\leq \cE\leq 3.41147.
\ee
In this way, we can plot the intervals of $\cE$ that corresponds to the branch cuts of $t(\cE,\hbar)$ for fixed $\hbar$.
We show it in Figure~\ref{fig:cuts-B3-1}.
In general, $F_{a/b}(\cE)$ is a degree-$b$ polynomial, and the energy conditions above
lead to $b$ intervals of $\cE$.
In our picture, the origin of $F_{a/b}(\cE)$ is the strong-weak relation \eqref{eq:strong-weak}.
Therefore this S-duality generates the complexity of the branch cut structure in Figure~\ref{fig:cuts-B3-1}.
Also, the figure is $4\pi$-periodic in $\hbar$, and it is also symmetric with respect to the vertical line $\hbar=2\pi$.
These properties are easily understood from the symmetric property of the K\"ahler modulus \eqref{eq:symmetry-t}. 

It turns out that the completely same figure appears as energy bands of electrons on a two-dimensional triangular lattice 
with a magnetic flux \cite{CW}.
This connection is simply expected by comparing the Hamiltonians of the two systems.
The tight-binding Hamiltonian of the electrons on the triangular lattice is given by
\be
\ba
H=T_x+T_x^\dagger+\lambda_1(T_y+T_y^\dagger)+\lambda_2( \re^{-\ri \phi/2} T_x T_y+\re^{\ri \phi/2} T_y^\dagger T_x^\dagger),
\ea
\label{eq:H-triangular}
\ee
where $T_x$ and $T_y$ are magnetic translation operators, which satisfy the commutation relation:
\be
T_x T_y=\re^{\ri \phi} T_y T_x,\qquad T_x T_x^\dagger=T_y T_y^\dagger=1.
\ee
Here $\phi$ is a magnetic flux through an elementary plaquette, and $\lambda_1$ and $\lambda_2$ are
anisotropy parameters.
If the magnetic flux is turned off, the spectrum is given by the following dispersion relation:
\be
\cE=2\cos k_x+2\lambda_1 \cos k_y+2\lambda_2 \cos (k_x+k_y).
\label{eq:disp}
\ee
Therefore in this case the spectrum is a single band.
The Hamiltonian \eqref{eq:H-triangular} for $\lambda_1=\lambda_2=1$ is very similar to the Hamiltonian \eqref{eq:H-B3} for $m_1=m_2=m_3=1$.
The big difference is that the former is periodic in $x$- and $y$-directions due to the lattice structure 
while the latter has no periodic structure for real $x$ and $y$.
This suggests us to perform an analytic continuation $x \to \ri x$ and $y \to \ri y$ in the latter.
Then the Hamiltonian \eqref{eq:H-B3}  for $m_1=m_2=m_3=1$ becomes
\be
H=\re^{\ri \x}+\re^{-\ri \x}+\re^{\ri \y}+\re^{-\ri \y}+\re^{\ri \hbar/2} \re^{\ri \x}\re^{\ri \y}+\re^{-\ri \hbar/2}\re^{-\ri \y}\re^{-\ri \x}.
\label{eq:H-B3-c}
\ee
The commutation relation is
\be
[\x, \y]=\ri \hbar \quad \Leftrightarrow \quad \re^{\ri \x}\re^{\ri \y}=\re^{-\ri \hbar} \re^{\ri \y}\re^{\ri \x}.
\ee
Now it is clear that the two Hamiltonians \eqref{eq:H-triangular} and \eqref{eq:H-B3-c} are identical under \eqref{eq:identification}.
Another way to see the similarity is to consider the Peiers--Onsager effective Hamiltonian,
which is essential obtained by ``quantizing'' the dispersion relation \eqref{eq:disp} by $[k_x, k_y]=\ri \phi$.
The spectrum of the Hamiltonian \eqref{eq:H-triangular} for $\lambda_1=\lambda_2=1$ was studied in \cite{CW},
and its band structure is in perfect agreement with Figure~\ref{fig:cuts-B3-1}.
Interestingly, this Hamiltonian is diagonalized by the so-called algebraic Bethe ansatz method \cite{WZ, FK}.
In this approach, the appearance of the quantum group structure is clear.

In the semiclassical regime $\phi \sim 0$, the width of each subband is extremely narrow.
As explained in \cite{HKT}, the position of each subband near $\cE=6$ is explained by the semiclassical expansion
\be
\cE_n^\text{weak}=6-\sqrt{3}(2n+1)\phi+\frac{2n^2+2n+1}{4}\phi^2-\frac{4n^3+6n^2+8n+3}{72\sqrt{3}} \phi^3+\cO(\phi^4).
\ee
This expansion is simply obtained by $\hbar \to -\phi$ in \eqref{eq:E-WKB}. This is a reflection of the identification \eqref{eq:identification}.

\begin{figure}[t]
\begin{center}
  \begin{minipage}[b]{0.45\linewidth}
    \centering
    \includegraphics[height=4cm]{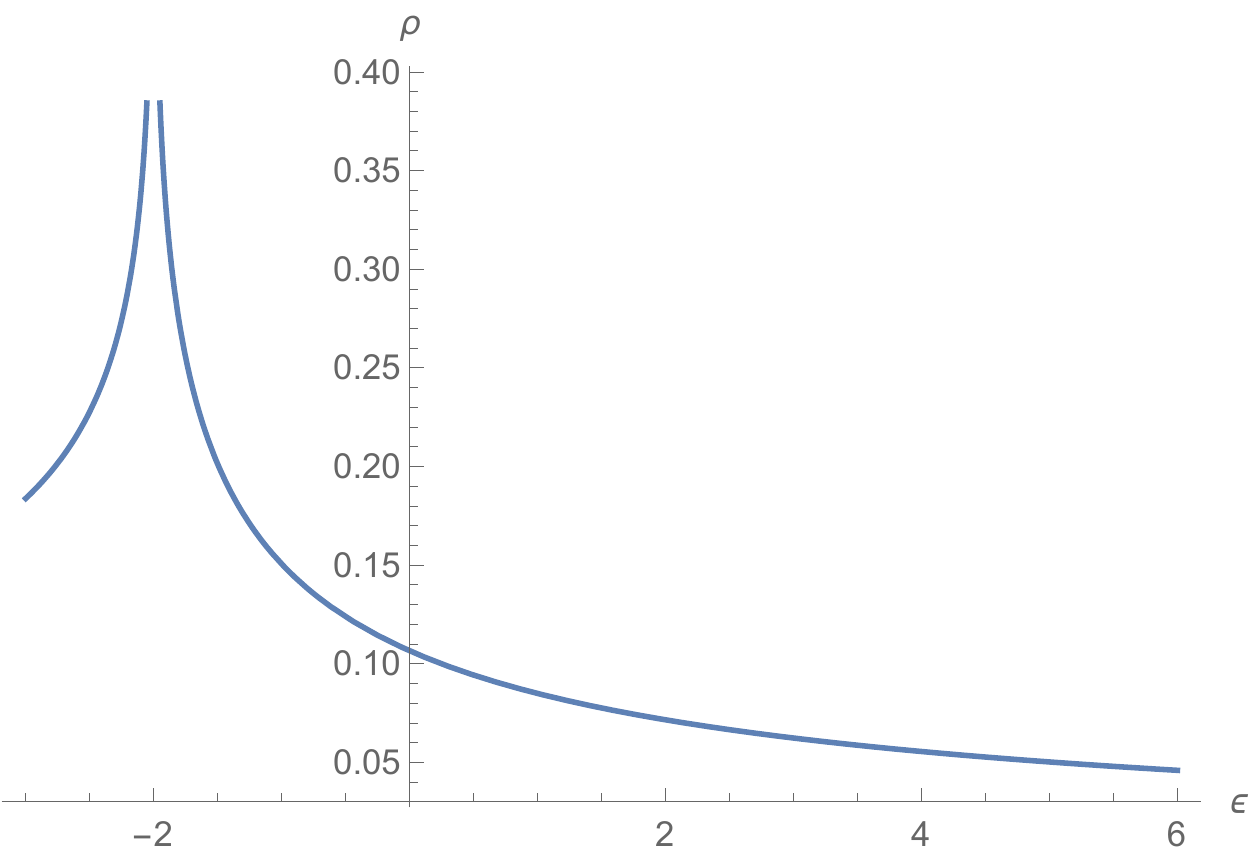}
  \end{minipage} 
  \begin{minipage}[b]{0.45\linewidth}
    \centering
    \includegraphics[height=4cm]{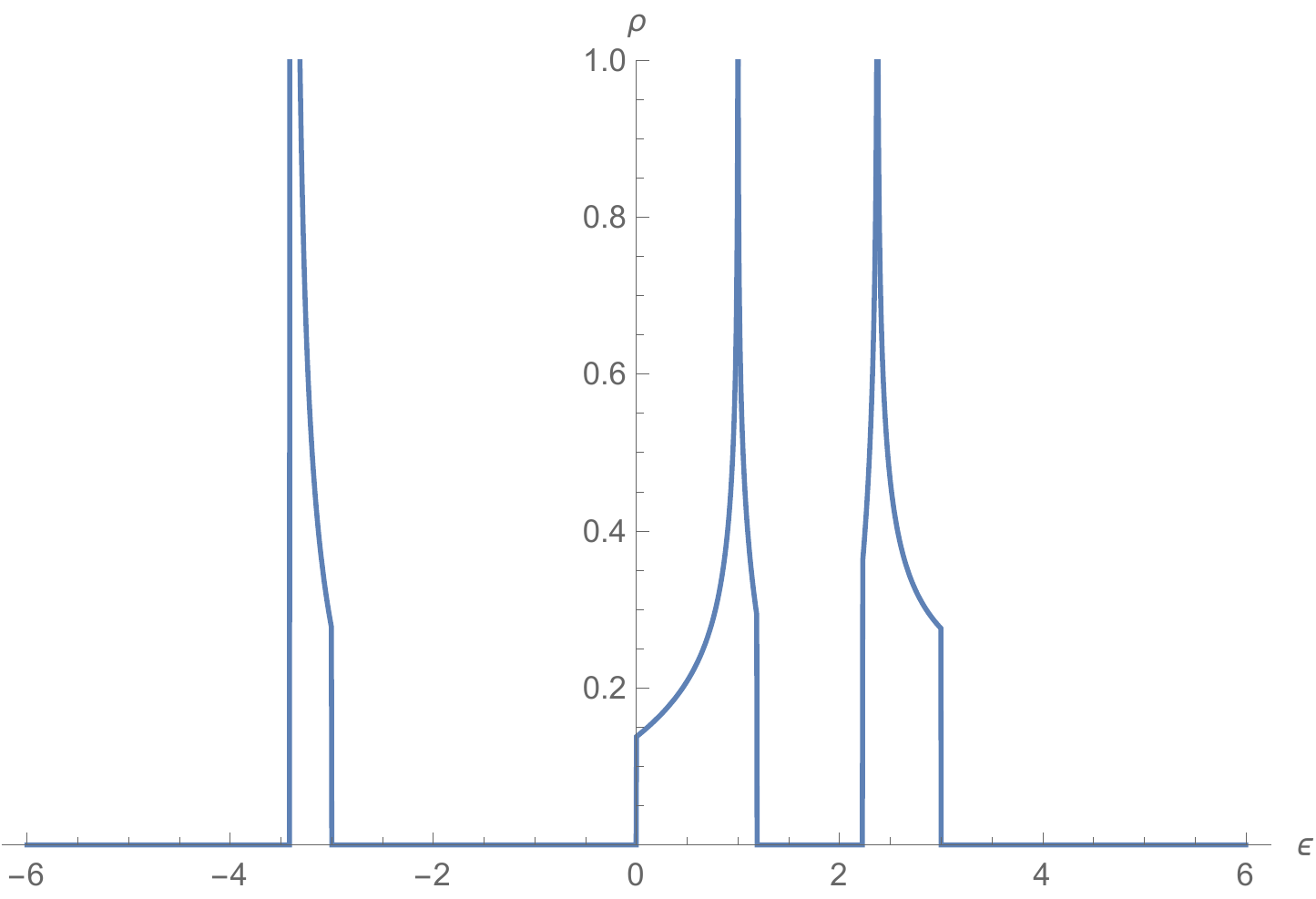}
  \end{minipage} 
\end{center}
  \caption{We show the density of states for the triangular lattice with $\lambda_1=\lambda_2=1$. 
The left figure is the graph for no flux ($\phi=0$), and the right is for $\phi=4\pi/3$.
The DOS is not vanishing only on the energy subbands, and it has singularities at $(-1)^{ab}F_{a/b}(\cE)=-2$.}
  \label{fig:DOS}
\end{figure}

Let us consider the physical interpretation of $t(\cE,\hbar)$.
As shown in \cite{HKT}, the derivative of $t(\cE,\hbar)$ w.r.t. $\cE$ is related to the density of states (DOS) of the electron system.
If the magnetic flux is turned off, the DOS in the current system for $\lambda_1=\lambda_2=1$ is computed by
\be
\rho_0(\cE)=\iint_0^{2\pi} \frac{\rd k_x \rd k_y}{4\pi^2}  \, \delta (2\cos k_x+2 \cos k_y+2 \cos (k_x+k_y)-\cE).
\ee
Performing the integrals, we find two expressions
\be
\rho_0(\cE)=\begin{cases}
\displaystyle \frac{1}{2\pi^2(3+\cE)^{1/4}} \eK \( \frac{12-\cE^2+8\sqrt{3+\cE}}{16\sqrt{3+\cE}} \), \quad & -2 < \cE \leq 6,  \vspace{0.2cm} \\
\displaystyle \frac{2}{\pi^2 \sqrt{12-\cE^2+8\sqrt{3+\cE}}} \eK \( \frac{16\sqrt{3+\cE}}{12-\cE^2+8\sqrt{3+\cE}} \), & -3 \leq \cE < -2,
\end{cases}
\ee
We show the behavior of this function in the left of Figure~\ref{fig:DOS}.
The DOS is singular at $\cE=-2$. This is known as the van Hove singularity.
Note that $\cE=-2$ is one of the zeros of the modular discriminant $\Delta=(\cE+2)^3(\cE+3)^2(\cE-6)$.
Therefore this point also corresponds to a regular singular point of the PF equation \eqref{eq:PF} or a singularity in the CY moduli space.

For $\phi=2\pi a/b$, the computation is more involved.
In this case, the dispersion relation is written as (see \cite{HK})
\be
F_{a/b}(\cE)=2\cos (bk_x')+2 \cos (b k_y)+2(-1)^{ab} \cos (b k_x'+ bk_y),
\ee
where the magnetic Brillouin zone is now defined by
\be
0 \leq k_x' \leq 2\pi/b, \qquad 0 \leq k_y \leq 2\pi.
\ee
Since $F_{a/b}(\cE)$ is the degree-$b$ polynomial, this defines the $b$ subbands.
Let us denote the dispersion relation on the $s$-th subband by
\be
\cE=\cE_s(k_x', k_y),\qquad s=1,\dots, b.
\ee 
Then, the DOS is computed by
\be
\rho(\cE, \phi=2\pi a/b)=\frac{b }{4\pi^2}\sum_{s=1}^b \int_0^{2\pi/b}\rd k_x' \int_0^{2\pi}\rd k_y   \, \delta (\cE_s(k_x', k_y)-\cE).
\ee
The evaluation of these integrals are essentially the same as the one for $\phi=0$ \cite{CW}, and
we finally obtain
\be
\rho(\cE,\phi=2\pi a/b)=\begin{cases}
\displaystyle \frac{|F'|}{2\pi^2 b(3+F)^{1/4}} \eK \( \frac{12-F^2+8\sqrt{3+F}}{16\sqrt{3+F}} \), \quad & -2 < F \leq 6,  \vspace{0.2cm} \\
\displaystyle \frac{2|F'|}{\pi^2 b\sqrt{12-F^2+8\sqrt{3+F}}} \eK \( \frac{16\sqrt{3+F}}{12-F^2+8\sqrt{3+F}} \), & -3 \leq F < -2,
\end{cases}
\ee
where $F=(-1)^{ab} F_{a/b}(\cE)$.
The DOS for $\phi=2\pi \cdot 2/3$ is shown in the right of Figure~\ref{fig:DOS}.

Let us compare the DOS with the quantum corrected K\"ahler modulus $t(\cE,\hbar)$.
The derivative is given by
\be
\frac{\pd t(\cE,\hbar=2\pi a/b)}{\pd \cE}= \frac{2F'}{\pi b \sqrt{F^2-12+8\sqrt{3+F}}} 
\eK \( \frac{16\sqrt{3+F}}{F^2-12+8\sqrt{3+F}} \).
\ee
For $F>6$, this function takes real values, but for $-3 \leq F \leq 6$, it has a non-vanishing imaginary part.
Using the identity of the elliptic integral
\be
\eK(1/z)=\sqrt{z}[ \eK(z)+\ri \eK(1-z) ],
\ee
one can analytically continue it to $-3 \leq F \leq 6$. 
After some computation, we find that the imaginary part agrees with the DOS above:
\be
\rho (\cE, \phi=2\pi a/b)=\frac{1}{\pi}\left|\im  \( \frac{\pd t(\cE, \hbar=2\pi a/b)}{\pd \cE}\)  \right|.
\ee

It is almost obvious that the anisotropy parameters $(\lambda_1, \lambda_2)$ are related to the mass parameters $(m_1, m_2, m_3)$.
To see this, we shift the variables $x \to x+\frac{1}{2}\log m_1$ and $y \to y+\frac{1}{2}\log m_2$.
Then the mirror curve \eqref{eq:mirror-B3} becomes
\be
\sqrt{m_1}(\re^{x}+\re^{-x})+\sqrt{m_2}(\re^{y}+\re^{-y})+\sqrt{m_1 m_2} m_3 \re^{x+y}+\frac{1}{\sqrt{m_1 m_2}} \re^{-x-y}=\cE.
\label{eq:mirror-B2-another}
\ee
If we require the additional condition
\be
m_1 m_2 m_3=1,
\ee
the coefficients of $\re^{x+y}$ and of $\re^{-x-y}$ agree.
Due to this additional constraint, free parameters are two of the three masses,
and they are related to $\lambda_1$ and $\lambda_2$. 
Dividing the both hand sides in \eqref{eq:mirror-B2-another} by $\sqrt{m_1}$, we get
\be
\re^{x}+\re^{-x}+\sqrt{\frac{m_2}{m_1}}(\re^{y}+\re^{-y})+\sqrt{\frac{m_3}{m_1}}(\re^{x+y}+\re^{-x-y})=\cE',
\ee
where $\cE'=\cE/\sqrt{m_1}$.
Comparing this curve with the dispersion relation \eqref{eq:disp}, it is natural to identify the parameters as
\be
\lambda_1=\sqrt{\frac{m_2}{m_1}}=m_2 \sqrt{m_3}, \qquad \lambda_2=\sqrt{\frac{m_3}{m_1}}=m_3 \sqrt{m_2}.
\ee
Under this identification, the quantum A-period \eqref{eq:qA-period} is rewritten as
\be
\ba
-t'&:=-t+\frac{1}{2} \log m_1=\log z'+(1+\lambda_1^2+\lambda_2^2){z'}^2+2(q^{1/2}+q^{-1/2})\lambda_1 \lambda_2 {z'}^3 \\
&\qquad+\biggl[ \frac{3}{2}(1+\lambda_1^4+\lambda_2^4)+(4+q+q^{-1})(\lambda_1^2+\lambda_2^2+\lambda_1^2\lambda_2^2 ) \biggr] {z'}^4+\cO({z'}^5).
\ea
\ee
where $z'=1/\cE'$.
The dual variables are defined by
\be
\wt{t}'=\wt{t}-\frac{1}{2} \log \wt{m}_1,\qquad
\wt{\lambda}_i=\lambda_i^{2\pi/\hbar}
\ee
Then, the S-duality relation \eqref{eq:t-Sdual} is mapped to the relation
\be
\wt{t}'(\wt{\cE}', \wt{\lambda}_i, \wt{\hbar} )=\frac{2\pi}{\hbar} t'(\cE', \lambda_i, \hbar).
\ee
\begin{figure}[t]
\begin{center}
  \begin{minipage}[b]{0.45\linewidth}
    \centering
    \includegraphics[width=0.95\linewidth]{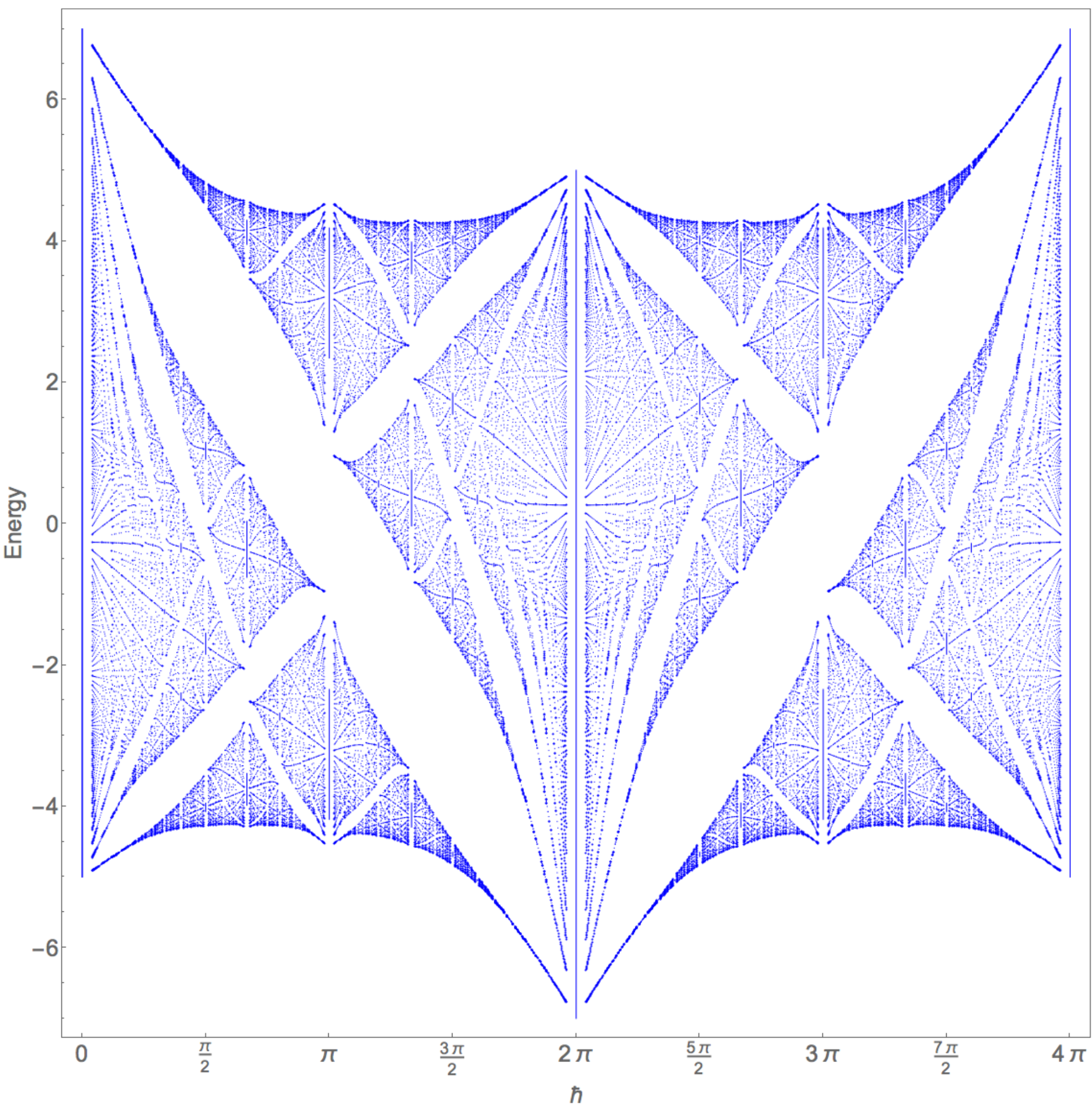} \\
   (a)
  \end{minipage}\hspace{0.5cm}
  \begin{minipage}[b]{0.45\linewidth}
    \centering
    \includegraphics[width=0.95\linewidth]{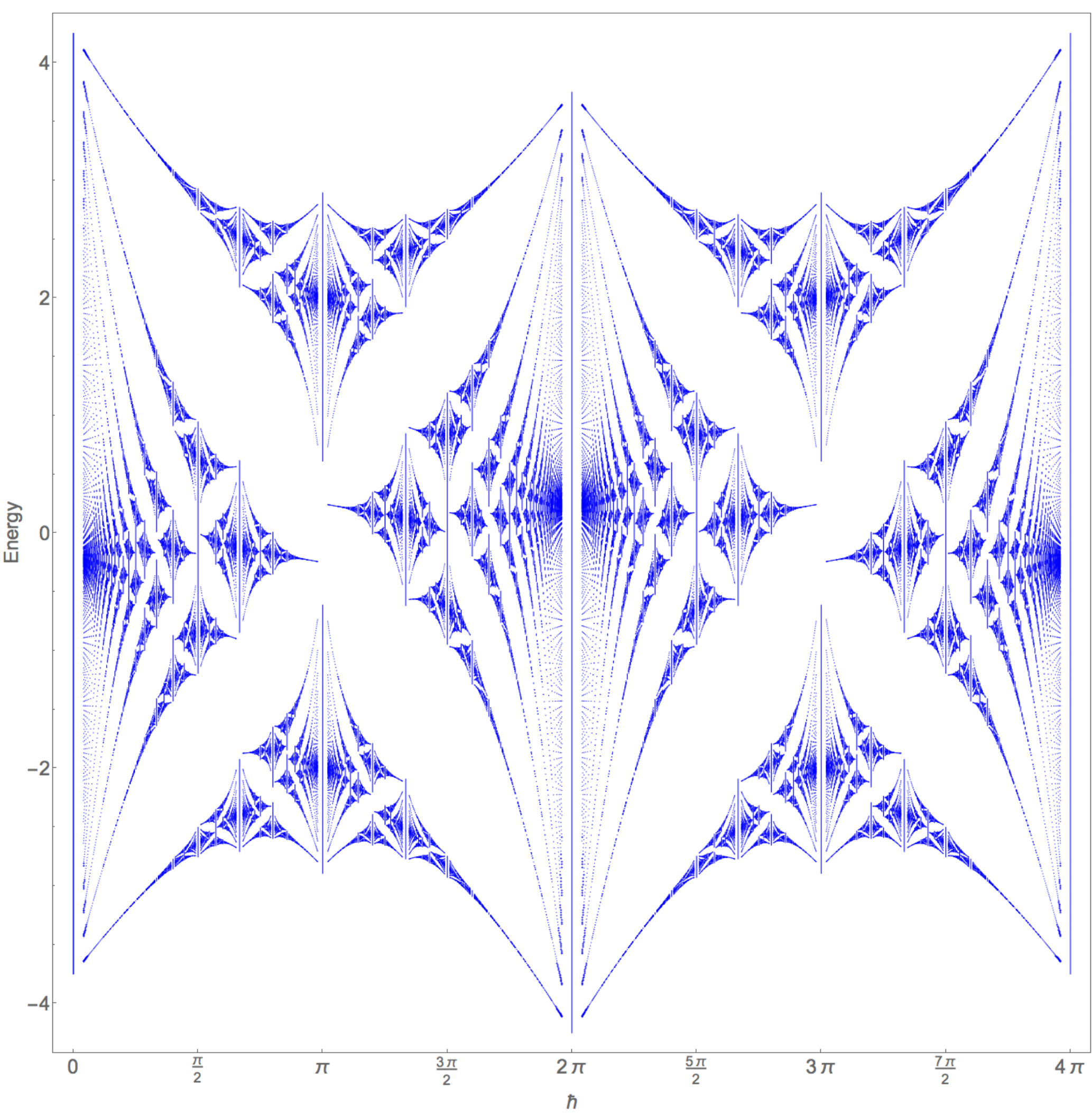} \\
    (b)
  \end{minipage} 
\end{center}
  \caption{The branch cut structure of the K\"ahler modulus $t'(\cE',\lambda_i, \hbar)$ for (a) $(\lambda_1,\lambda_2)=(2,1/2)$ and (b)
 $(\lambda_1,\lambda_2)=(1,1/8)$.}
  \label{fig:cuts-B3-2}
\end{figure}
This relates the two rescaled energies $\cE'$ and $\wt{\cE}'$.
The explicit relation can be written down by using the result in \cite{HK}, as in the case of $\lambda_1=\lambda_2=1$ in the previous subsection.
Since the calculation is straightforward, we do not write it down here.
The branch cut structure of the K\"ahler modulus $t'$ is also determined by the same way above.
We show it in Figure~\ref{fig:cuts-B3-2}.
In the limit $\lambda_2 \to 0$, the problem is reduced to the original problem of Hofstadter for electrons on the square lattice.
One can see that the right of Figure~\ref{fig:cuts-B3-2} for $\lambda_1=1$ and $\lambda_2=1/8$ is actually similar to the Hofstadter butterfly in \cite{Hof}.

\section{Concluding remarks}\label{sec:conclusion}
In this paper, we further explored a connection between quantum geometry for toric Calabi--Yau threefolds
and 2d electron systems, recently pointed out in \cite{HKT}.
We found the precise correspondence that the local $\cB_3$ geometry is associated with
electrons on the triangular lattice. This correspondece helps us to study the analytic property
in quantum geometry of this toric CY from known results in condensed matter physics.
In fact, the strong-weak coupling relation \eqref{eq:strong-weak} for the energies in the quantum mechanical system for local $\cB_3$
was conjectured by using the result on the triangular lattice in \cite{HK}.
Based on this result, we identified the branch cut structure of the K\"ahler modulus,
and its figure turned out to be equivalent to the band spectrum of the tight-binding Hamiltonian on the triangular lattice.
The K\"ahler modulus has a direct physical interpretation as the density of states in the electron system.
In this sense, the quantum A-period of local $\cB_3$ knows
all the spectral information of the electrons on the triangular lattice.

\begin{figure}[t]
\begin{center}
  \begin{minipage}[b]{0.4\linewidth}
    \centering
    \includegraphics[width=6cm]{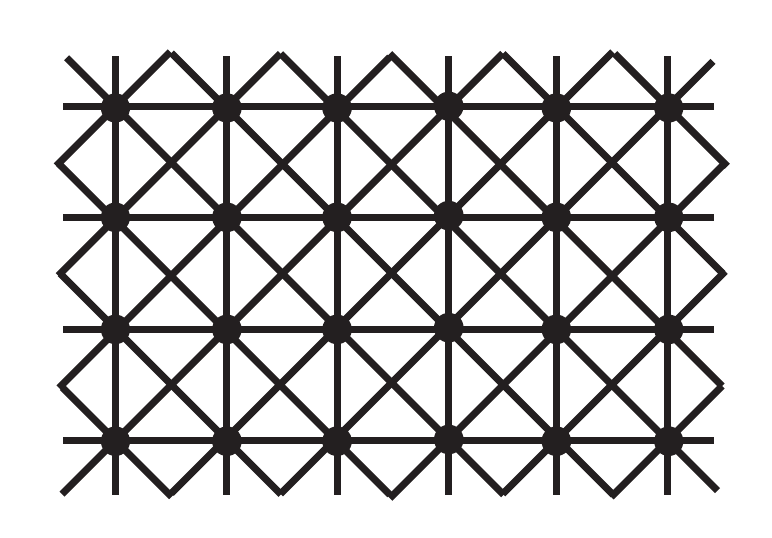}
  \end{minipage} \hspace{1cm}
  \begin{minipage}[b]{0.4\linewidth}
    \centering
    \includegraphics[width=4cm]{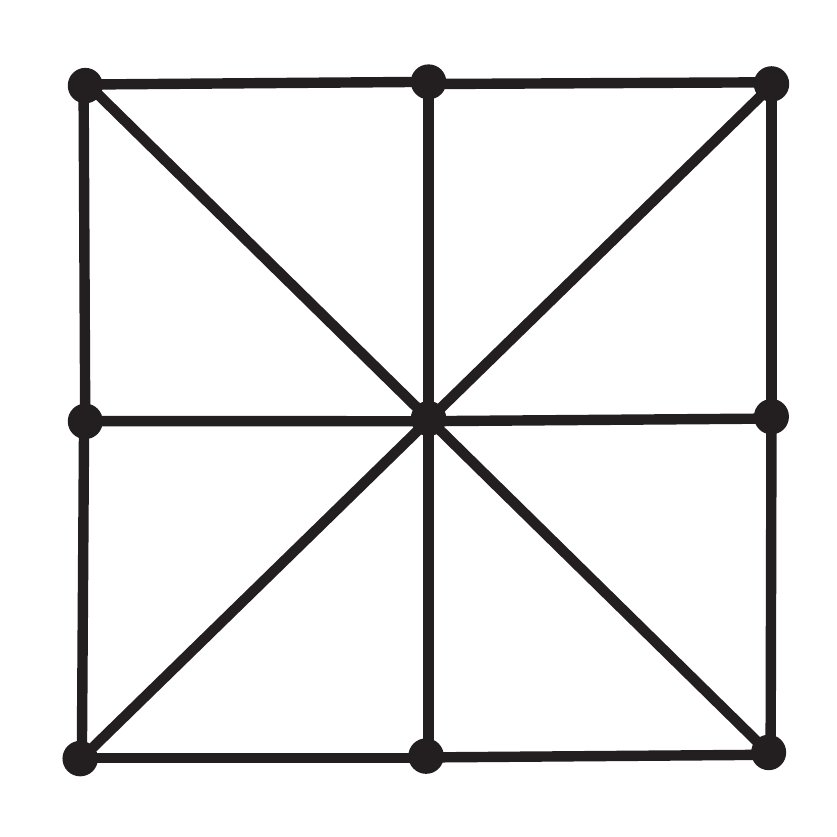}
  \end{minipage} 
\end{center}
  \caption{Left: The 2d lattice considered in \cite{HK}. An electron can hop to eight possible sites connected by the solid lines.
  Right: The toric diagram of the corresponding CY threefold.}
  \label{fig:P15}
\end{figure}

As further generalization, we remark two directions.
In \cite{HK}, the authors considered a more complicated system, where there is an additional next-nearest-neighbor hopping interaction, 
shown in the left of Figure~\ref{fig:P15}.
The Hamiltonian of this system reads
\be
H=T_x+T_x^\dagger+\lambda_1(T_y+T_y^\dagger)+\lambda_2( \re^{-\ri \phi/2} T_x T_y+\re^{\ri \phi/2} T_y^\dagger T_x^\dagger)
+\lambda_3( \re^{\ri \phi/2} T_x T_y^\dagger+\re^{-\ri \phi/2} T_y T_x^\dagger).
\ee
Of course, if the hopping parameter $\lambda_3$ is turned off, it reduces to the Hamiltonian for the triangular lattice.
This motivates us to consider the eigenvalue problem of the related Hamiltonian:
\be
H=\re^{\x}+\re^{-\x}+\lambda_1( \re^{\y}+\re^{-\y})+\lambda_2(\re^{\x+\y}+\re^{-\x-\y})
+\lambda_3(\re^{\x-\y}+\re^{-\x+\y}).
\label{eq:H-P15}
\ee
This Hamiltonian is also associated with the mirror geometry for a toric Calabi--Yau threefold, whose toric diagram is 
shown in the right of Figure~\ref{fig:P15}. The geometry is sometimes called 
local $\cB_5$ in the literature.
Therefore it is expected that the quantum Hamiltonian is also diagonalized by the refined topological string in this CY geometry.
It would be interesting to confirm the relation of these two systems in more detail along the line of this paper.

\begin{figure}[t]
\begin{center}
    \includegraphics[width=.55\textwidth]{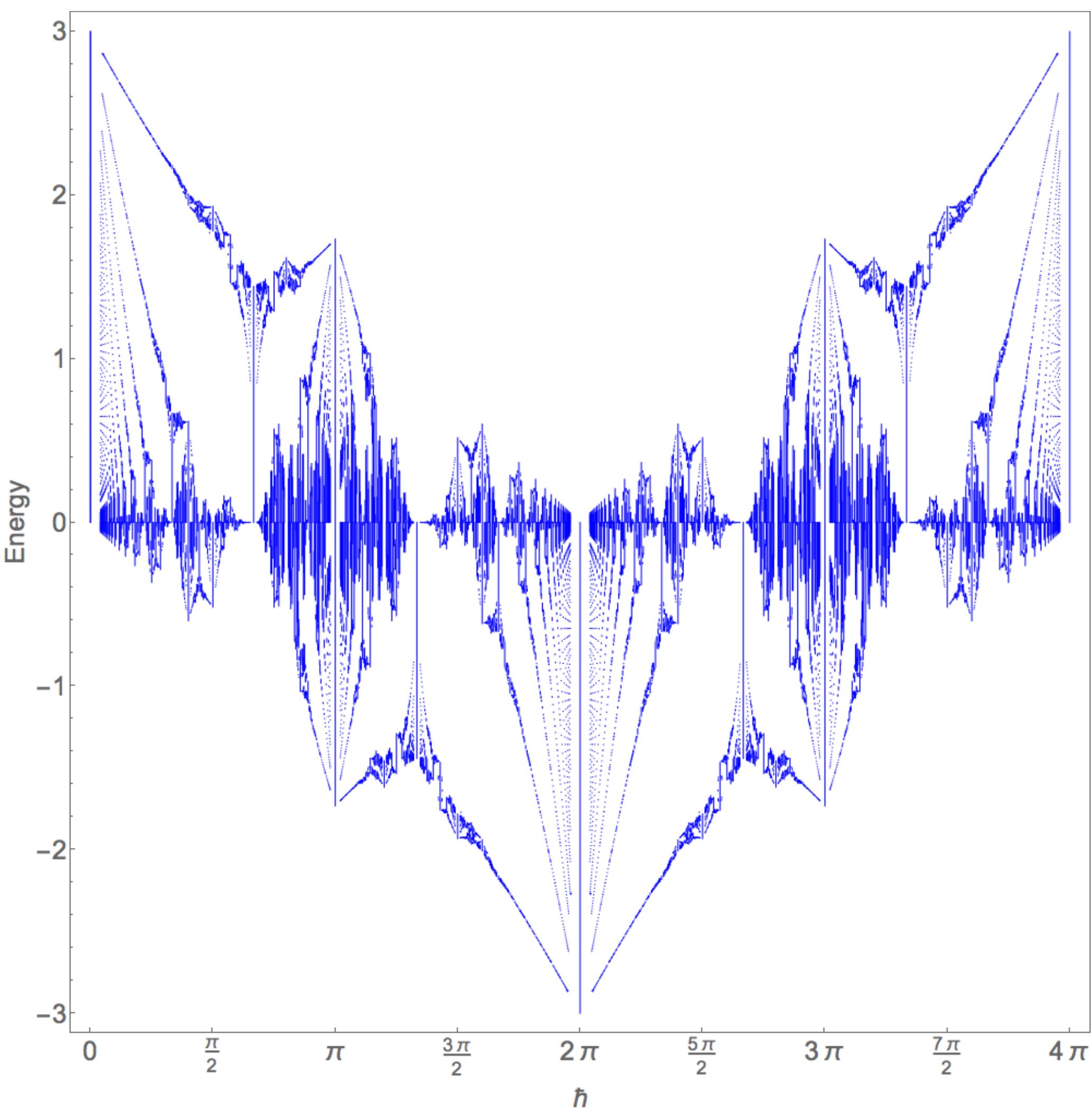}
\end{center}
\caption{The cut structure for local $\mathbb{P}^2$.}
\label{fig:local-P2}
\end{figure}

We also encounter another puzzling problem if considering a much simpler geometry, called local $\mathbb{P}^2$.
The Hamiltonian of this geometry is given by
\be
H=\re^{\x}+\re^{\y}+\re^{-\x-\y}.
\ee
On one hand, the analogy with local $\cB_3$ leads to the following Hamiltonian
\be
H=T_x+T_y+\re^{\ri \phi/2} T_y^\dagger T_x^\dagger.
\ee
However, this Hamiltonian is obviously non-Hermitian, and it seems to have no physical interpretation.
On the other hand, by the same procedure in the previous section, we can still determine the branch cut structure
of the K\"ahler modulus for local $\mathbb{P}^2$.
We show it in Figure~\ref{fig:local-P2}.
Is this figure identified as a band spectrum of certain electron system?
We do not have any answers for this question, and it would be interesting to find a resolution of this problem.
If this problem is solved, one can easily generalize it to a wide class of toric CYs. 

\acknowledgments{
We thank Santiago Codesido, Jie Gu, Hosho Katsura, Marcos Mari\~{n}o, Kazumi Okuyama, Yuji Tachikawa, Futoshi Yagi and Szabolcs Zakany 
for useful discussions.
We are grateful to Marcos Mari\~no and Yuji Tachikawa for giving us helpful comments on the manuscript.
The work of YH is supported in part by the Fonds National Suisse, subsidies 200021-156995 and by the NCCR 51NF40-141869 
``The Mathematics of Physics'' (SwissMAP).
}

\appendix

\section{Quantum periods and refined topological vertex}\label{sec:RTV}
To solve the exact quantization condition, we need the NS free energy and the quantum A-period (or the quantum mirror map).
The former can be computed by using the refined topological vertex formalism \cite{AKMV, IKV}.
The latter is obtained from the quantum mirror curve \cite{ACDKV}.%
\footnote{In principle, one can also compute the quantum B-period from the mirror curve, as explained in \cite{ACDKV}.
However it is much more involved than the computation of the A-period.
Instead, we here use the topological vertex to compute the B-period. These two different methods give the same result
thanks to mirror symmetry.}  
We here review these methods quickly.

\subsection{Quantum A-period}
The quantization of the mirror curve \eqref{eq:mirror-B3} leads to the difference equation \eqref{eq:diff-eq}.
Introducing a ratio function
\be
V(X)=\frac{\psi(x-\ri \hbar)}{\psi(x)}, \qquad X=\re^{x},
\ee
we obtain the following $q$-difference equation
\be
X+\frac{m_1}{X}+V(X)+\frac{m_2}{V(q X)}+\frac{q^{-1/2}}{X V(q X)}+m_3 q^{-1/2} X V(X)=\frac{1}{z}.
\label{eq:q-diff}
\ee
We solve this equation order by order in $z$:
\be
V(X)=\frac{v_{-1}(X)}{z}+v_0(X)+v_1(X)z+\cdots.
\ee
It is easy to fix each coefficient:
\be
v_{-1}(X)=\frac{1}{1+m_3 q^{-1/2}X}, \quad
v_0(X)=-\frac{m_1+q^{-1/2}X^2}{X(1+m_3 q^{-1/2} X)}.
\ee
Then,
\be
\log V(X)=\log\( \frac{v_{-1}(X)}{z} \)+\frac{v_0(X)}{v_{-1}(X)}z+\cdots.
\ee
The quantum A-period is obtained by the formula
\be
\Pi_A(z;q)=-\Res_{X=0}  \frac{V(X)-\log(v_{-1}(X)/z)}{X} 
=- \Res_{X=0} \(  \frac{v_0(X)}{X v_{-1}(X)}z+\cdots\).
\ee
Using the explicit solution to \eqref{eq:q-diff}, we finally obtain \eqref{eq:qA-period}.
In the classical limit $q \to 1$, it reduces to the classical A-period \eqref{eq:A-period}.

\subsection{NS free energy from refined topological vertex}
In this appendix, we present how to calculate the Nekrasov--Shatashvili free energy. This free energy can be obtained by taking the Nekrasov--Shatashvili limit for the refined topological string partition function as in \eqref{eq:F-NS}. The refined topological string free energy is defined by using the refined topological string partition function as 
\be
F_{\text{ref}}(t;q_1,q_2)=-\log Z_\text{ref} (t;q_1 , q_2),
\ee
where $q_{1}$ and $q_{2}$ are related to the refined topological string couplings $\epsilon_1$ and $\epsilon_2$, 
\be
q_1=\mathrm{e}^{\epsilon_1},\qquad q_2=\mathrm{e}^{-\epsilon_2}.
\ee
Note that for $q_{1}=q_{2}$ this free energy reduces to the usual topological string. 
Therefore the partition function $Z_\text{ref} (t;q_1 , q_2)$ is a one-parameter deformation of the usual topological string partition function.
This partition function can be obtained from the refined topological vertex formalism \cite{IKV}, reviewed here. 
The refined topological vertex $C_{\lambda \mu \nu}(q_1,q_2)$ is defined by,%
\footnote{The notation of the refined topological vertex is different from the usual one. However, in order to avoid confusing the K\"ahler moduli with the refined topological string coupling, we use this notation.}
\be
\ba
C_{\lambda \mu \nu}(q_1,q_2) &= q_1^{-\frac{||\mu^{t}||^{2}}{2}}q_2^{\frac{||\mu||^2 + ||\nu||^{2}}{2}} \tilde{Z}_{\nu}(q_1,q_2)
 \sum_{\eta}\Bigl(\frac{q_2}{q_1}\Bigr)^{\frac{|\eta| + |\lambda| - |\mu|}{2}}  s_{\lambda^{t}/\eta}(q_1^{-\rho}q_2^{-\nu})s_{\mu/\eta}(q_1^{-\nu^{t}}q_2^{-\rho}), \\
\tilde{Z}_{\nu}(q_1,q_2) &= \prod_{(i,j) \in \nu}(1-q_2^{\nu_{i}-j}q_1^{\nu_{j}^{t} -i +1})^{-1},
\ea
\ee
where $|\mu|$ is the total number of boxes in the Young diagram $\mu$, $s_{\mu/\nu}(x)$ is the skew-Schur function, and $||\mu||$ and $\rho$ are defined as follows:
\be
||\mu||:=\sum_{i=1}^{l(\mu)}\mu^{2}_{i}, \hspace{2em}
\rho:=-i+\frac{1}{2}.
\ee
\begin{figure}[tb]
\centering
    \includegraphics[width=5cm]{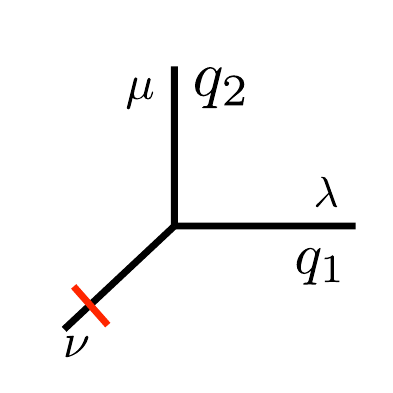}
    \caption{The pictorial description of the refined topological vertex. The red line is the preferred direction.}
    \label{vertex}
\end{figure}
\par
\begin{figure}[t]
\begin{center}
  \begin{minipage}[b]{0.45\linewidth}
    \centering
    \includegraphics[width=5cm]{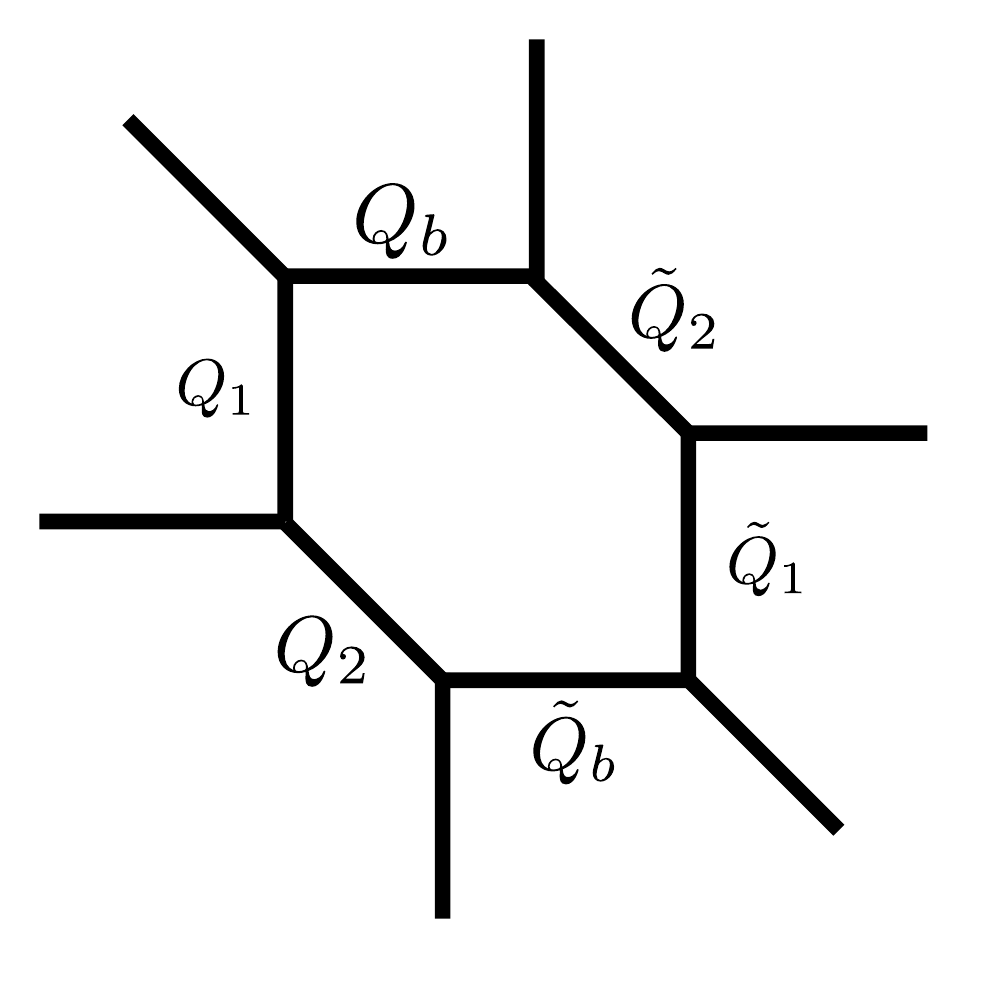} 
  \end{minipage}
  \begin{minipage}[b]{0.45\linewidth}
    \centering
    \includegraphics[width=6cm]{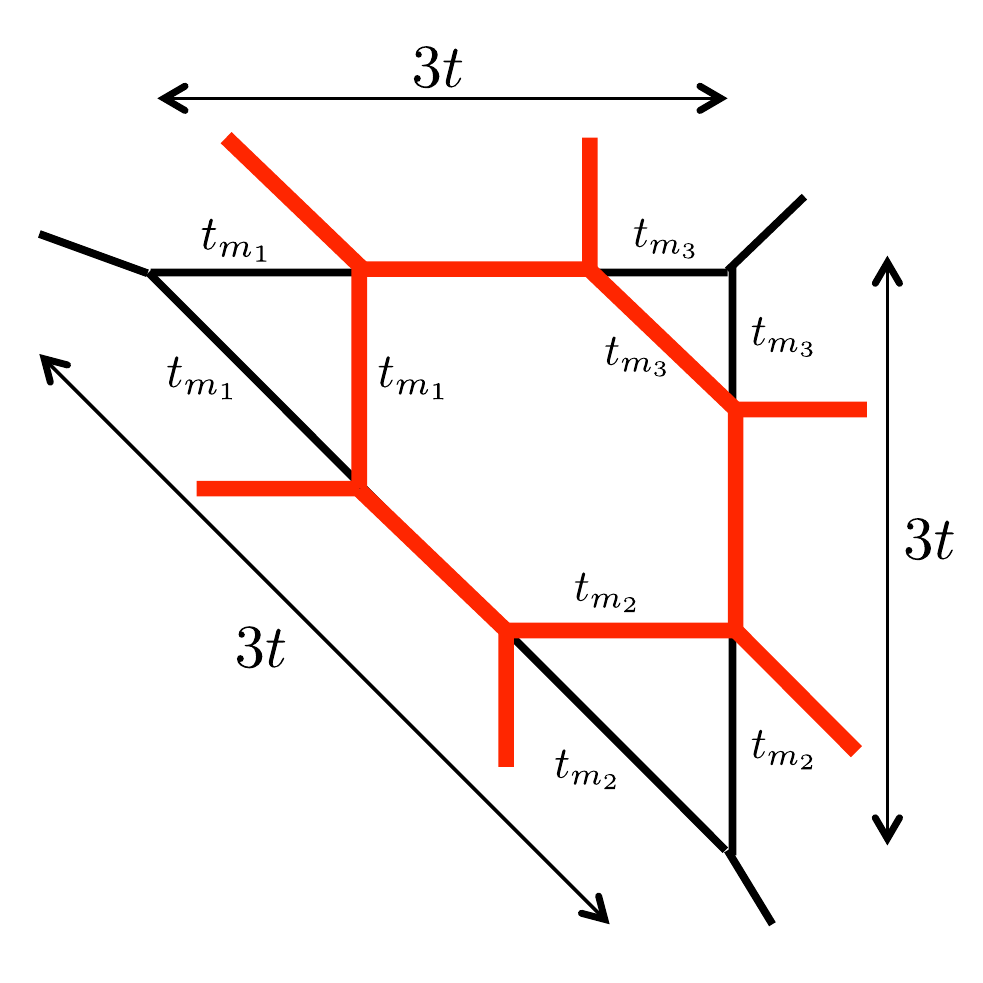}
  \end{minipage} 
\end{center}
  \caption{Left: The web diagram of local $\mathcal{B}_{3}$. Right: We can embed it into the web diagram of local $\mathbb{P}^2$.}
  \label{web}
\end{figure}

Using this building block, we compute the refined free energy for local $\cB_3$.
Note that the refined free energy for the same geometry but in a slightly different convention was computed in \cite{Sugimoto}.
The web diagram of local $\cB_{3}$ is shown in the left of Figure~\ref{web}, 
and we can write the instanton part of the partition function $Z_\text{ref}^\text{inst}$ 
in local $\cB_{3}$  in terms of the refined topological vertex,

\be
\ba
Z_\text{ref}^\text{inst}
=
\sum_{\text{all indices}}
&(-Q_{1})^{|\mu_{1}|}(-\tilde{Q}_{1})^{|\tilde{\mu}_{1}|}
(-Q_{2})^{|\mu_{2}|}(-\tilde{Q}_{2})^{|\tilde{\mu}_{2}|}
(-Q_{b})^{|\mu_{b}|}(-\tilde{Q}_{b})^{|\tilde{\mu}_{b}|}
 \\
&\times
C_{\emptyset \mu_{1} \mu_{b}}(q_1,q_2)C_{\mu_{2}\mu^{t}_{1}\emptyset}(q_2,q_1)C_{\mu^{t}_{2}\emptyset\tilde{\mu}_{b}}(q_1,q_2)
\\
&\times
C_{\tilde{\mu}_{2}\emptyset\mu^{t}_{b}}(q_2,q_1)C_{\tilde{\mu}_{2}\tilde{\mu}_{1}\emptyset}(q_1,q_2)C_{\emptyset\tilde{\mu}_{1}\tilde{\mu}^{t}_{b}}(q_2,q_1).
\ea
\ee
After some calculation, we obtain
\be
\ba
Z_\text{ref}^\text{inst}
=
\sum_{\text{all indices}}
&\tilde{Z}_{\mu_{b}}(q_1,q_2)\tilde{Z}_{\tilde{\mu}_{b}}(q_1,q_2)
\tilde{Z}_{\mu^{t}_{b}}(q_2,q_1)\tilde{Z}_{\tilde{\mu}^{t}_{b}}(q_2,q_1)
q_1^{\frac{||\mu^{t}_{b}||^2 + ||\tilde{\mu}^{t}_{b}||^2}{2}}
q_2^{\frac{||\mu_{b}||^2 + ||\tilde{\mu}_{b}||^2}{2}}
 \\
&\times
\prod_{i,j=1}^{\infty}
\frac{
(1-Q_{1}q_1^{-\mu^{t}_{b,j}+i-\frac{1}{2}}q_2^{j-\frac{1}{2}})
(1-Q_{2}q_1^{i-\frac{1}{2}}q_2^{-\tilde{\mu}_{b,i}+j-\frac{1}{2}})
}{
(1-Q_{1}Q_{2}q_1^{-\mu^{t}_{b,j}+i}q_2^{-\tilde{\mu}_{b,i}+j-1})
}
\\
&\times
\prod_{i,j=1}^{\infty}
\frac{
(1-\tilde{Q}_{1}q_1^{i-\frac{1}{2}}q_2^{-\tilde{\mu}_{b,i}+j-\frac{1}{2}})
(1-\tilde{Q}_{2}q_1^{-\mu^{t}_{b,j}+i-\frac{1}{2}}q_2^{j-\frac{1}{2}})
}{
(1-\tilde{Q}_{1}\tilde{Q}_{2}q_1^{-\mu^{t}_{b,j}+i-1}q_2^{-\tilde{\mu}_{b,i}+j})
}
\ea
\ee
where we used some formula for the skew-Schur function,
\be
\ba
s_{\lambda/\mu}(\alpha \bold{x}) &= \alpha^{|\lambda|-|\mu|}s_{\lambda/\mu}(\bold{x}),
 \\
\sum_{\eta}s_{\eta/\lambda}(\bold{x})s_{\eta/\mu}(\bold{y}) &= \prod_{i,j=1}^{\infty}(1-x_{i}y_{j})^{-1}\sum_{\tau}s_{\mu/\tau}(\bold{x})s_{\lambda/\tau}(\bold{y}),
 \\
\sum_{\eta}s_{\eta^{t}/\lambda}(\bold{x})s_{\eta/\mu}(\bold{y}) &= \prod_{i,j=1}^{\infty}(1+x_{i}y_{j})\sum_{\tau}s_{\mu^{t}/\tau}(\bold{x})s_{\lambda^{t}/\tau^{t}}(\bold{y}).
\label{schur}
\ea
\ee
Due to the shape of the web diagram, we have following constraints,
\be
\ba
Q_1 Q_2=\tilde{Q}_1 \tilde{Q}_2 , \qquad Q_b \tilde{Q}_2 = \tilde{Q}_b Q_1 .
\ea
\label{eq:Q-rel}
\ee
For our purpose, we need to know the relation between the K\"ahler moduli $(Q_i$, $\tilde{Q}_i)$
and the mass parameters $m_i$ in the mirror curve.
This can be done as follows.
Since local $\cB_3$ is a three-point blow-up of local $\mathbb{P}^2$, 
we can ``embed'' its web diagram into that of local $\mathbb{P}^2$ as shown in the right of Figure~\ref{web}.
The parameters $t_{m_i}$ in this figure is directly related to the mass parameters as
\be
t_{m_i}=t+\log m_i, \qquad i=1,2,3.
\ee
Then, one finds
\be
\ba
Q_1&=\re^{-t_{m_1}}=\frac{Q}{m_1}, \\
Q_2&=\re^{-(3t-t_{m_1}-t_{m_2})}=m_1 m_2 Q,\\
Q_b&=\re^{-(3t-t_{m_1}-t_{m_3})}=m_1 m_3 Q.
\ea
\ee
where $Q=\re^{-t}$ is the ``true modulus'' in local $\cB_3$.
Similarly, one obtains
\be
\ba
\tilde{Q}_1&=m_2 m_3 Q,\qquad \tilde{Q}_2=\frac{Q}{m_3}, \qquad \tilde{Q}_b =\frac{Q}{m_2},
\ea
\ee
Of course, these satisfy the relations \eqref{eq:Q-rel}.
Under this identification, 
the refined free energy up to $\cO(Q^2)$  is explicitly given by
\be
\ba
F_\text{ref}^\text{inst}(t;q_1,q_2)
=&-\frac{\sqrt{q_1 q_2}}{(q_1-1) (q_2-1)}
\biggl(m_1 m_2 + m_2 m_3 + m_3 m_1 +\frac{1}{m_1}+\frac{1}{m_2}+\frac{1}{m_3}
\biggr)Q
\\
&+\frac{1}{(q_1^2-1)(	q_2^2-1)}
\biggl[
-(q_1+q_2+q_1^2+q_2^2+2q_1 q_2(q_1+q_2))(m_1 +m_2 +m_3)
\\
&+\frac{q_1 q_2}{2}\biggl(m_1^2 m_2^2 + m_2^2 m_3^2 + m_3^2 m_1^2 +\frac{1}{m_1^2}+\frac{1}{m_2^2}+\frac{1}{m_3^2}\biggr)
\biggr]Q^2 + \cO(Q^3).
\ea
\label{eq:Fref-2inst}
\ee
Taking the NS limit, one finally obtains the NS free energy \eqref{eq:F-NS-B3}.

\section{Spectral functions}\label{sec:spectral}
In the main text, we focus on the quantization condition that determines the exact eigenvalues of the quantum mirror curve.
There are two interesting quantities associated with the spectrum.
One is the spectral zeta function $\zeta_S(s)$, and the other is the spectral determinant $D(\cE)$.
These are defined by
\be
\zeta_S(s):= \sum_{n=0}^\infty \frac{1}{\cE_n^s},\qquad
D(\cE):=\prod_{n=0}^\infty \(1-\frac{\cE}{\cE_n} \).
\ee
For the eigenvalues $\cE_n$ of the quantum mirror curve, the sum of the spectral zeta function is convergent for $\real s>0$.%
\footnote{The fact that the sum of $\zeta_S(s)$ converges for any positive integer $s$ means that the operator $H^{-1}$ is
a trace-class operator \cite{KM}. In particular, $\zeta_S(1)=\Tr (H^{-1})$ is well-defined, and we do not need any regularization
to evaluate it. This is different from the spectral zeta function in the harmonic oscillator.}
To extend it to $\real s < 0$, one needs an analytic continuation.
These two functions are related by
\be
\log D(\cE)=-\sum_{\ell=1}^\infty \frac{\zeta_S(\ell)}{\ell} \cE^\ell.
\ee
Obviously, the spectral determinant vanishes only for $\cE=\cE_n$ ($n=0,1,2,\dots$).
This means that if we know this function, then we can determine all the eigenvalues as its zeros.
This idea was used in \cite{GHM1}.
There, the exact form of the spectral determinant was conjectured in terms of the topological string free energy.
The exact quantization condition was determined as the zero locus of this function.
The resulting quantization condition looks very complicated, but in \cite{WZH} it turned out to be written
as the simple form \eqref{eq:EQC}.
The interesting point is that to construct the spectral determinant, one needs the usual (or unrefined) topological string free energy
as well as the NS free energy.
On the other hand, we have already seen that the exact quantization condition \eqref{eq:EQC} consists of
only the NS free energy \cite{WZH}.
This implies that there is a non-trivial relation between the unrefined free energy and the NS free energy.
This connection was recently studied for several toric CYs in \cite{SWH, GG}.

Here we summarize the result in \cite{GHM1}.
For simplicity, we restrict ourselves to the case of $m_1=m_2=m_3=1$.
It is convenient to change the notation slightly,
\be
\Xi(\mu) := D(-\re^{\mu})=\prod_{n=0}^\infty (1+\re^{\mu-E_n}),
\ee
where $E_n=\log \cE_n$.
This function is regarded as the grand canonical partition function for an ideal Fermi-gas with the one-particle energy $E_n$.
The parameter $\mu$ plays the role of the chemical potential in this picture.
The conjecture in \cite{GHM1} states that the spectral determinant is written as an infinite sum
\be
\Xi(\mu)=\sum_{m \in \mathbb{Z}} \re^{J(\mu+2\pi \ri m)}.
\ee
The building block $J(\mu)$ is related to the refined topological string free energy.
The large $\mu$ expansion of $J(\mu)$ consists of the following two non-trivial parts:
\be
J(\mu)=\frac{T^3}{2\pi \hbar}-\frac{\hbar}{8\pi}T+A(\hbar)+J_\text{NS}(\mu)+J_\text{top}(\mu).
\label{eq:J}
\ee
where the parameter $T$ is related to the K\"ahler modulus $t$ by
\be
t=T+\pi \ri,
\ee
while the chemical potential is related to the energy by
\be
\cE=-\re^{\mu} \quad \Leftrightarrow \quad z=-\re^{-\mu}.
\ee
In the end, the relation between $T$ and $\mu$ is the (sign-reversed) quantum A-period:
\be
T=\mu-3\re^{-2\mu}+2(q^{1/2}+q^{-1/2})\re^{-3\mu}-3\( \frac{11}{2}+q+q^{-1} \)\re^{-4\mu}+\cO(\re^{-5\mu}).
\ee
where we used the quantum A-period \eqref{eq:qA-period}.
The two ingredients $J_\text{NS}(\mu)$ and $J_\text{top}(\mu)$ are directly related to the NS free energy and the unrefined free energy, respectively:
\be
\ba
J_\text{NS}(\mu)&=\frac{T}{2\pi}\frac{\pd}{\pd T}F_\text{NS}^\text{inst}(T+\pi \ri, \hbar)+\frac{\hbar^2}{2\pi}\frac{\pd}{\pd \hbar}
\( \frac{F_\text{NS}^\text{inst}(T+\pi \ri,\hbar)}{\hbar} \), \\
J_\text{top}(\mu)&=-F_\text{top}^\text{inst} \( \frac{2\pi T}{\hbar}, \frac{4\pi^2}{\hbar} \).
\ea
\ee
where the unrefined topological string free energy is obtained from the refined topological string energy by
\be
F_\text{top}(t, g_s) = \lim_{q_2 \to q_1} F_\text{ref}(t; q_1, q_2),\qquad q_1=\re^{\ri g_s}.
\ee
Note that our convention of the refined free energy \eqref{eq:Fref-2inst} is slightly different from the one in \cite{GHM1}.

One interesting consequence of \eqref{eq:J} is that if one considers the double scaling limit
\be
\hbar \to \infty, \qquad T \to \infty, \qquad \wt{T}=\frac{2\pi T}{\hbar}\;\; \text{kept finite},
\ee
then $J_\text{top}^\text{inst}(\mu)$ survives. The NS part $J_\text{NS}^\text{inst}(\mu)$ is sufficiently suppressed in this limit.
Therefore this 't Hooft-like limit allows us to extract the information on the unrefined topological string free energy.
The sufficiently small correction $J_\text{NS}^\text{inst}(\mu)$ is then regarded as a nonperturbative correction
to the unrefined topological string.
This perspective was studied in \cite{MZ, KMZ} in detail.
Note that the exact quantization condition \eqref{eq:EQC} does not have an obvious relation to the unrefined free energy.
We need to consider the spectral determinant to know about it.

\section{Relation to mass deformed $E_8$ del Pezzo geometry}\label{sec:E8}
In subsection~\ref{subsec:classical}, we remarked that the Picard--Fuchs equation \eqref{eq:PF} for $m_1=m_2=m_3=1$
is identical to the one for the mass deformed $E_8$ del Pezzo geometry for a particular choice of mass parameters.
As we will see in this appendix, this relation turns out to hold more widely.

\begin{figure}[t]
\begin{center}
  \begin{minipage}[b]{0.45\linewidth}
    \centering
    \includegraphics[height=4cm]{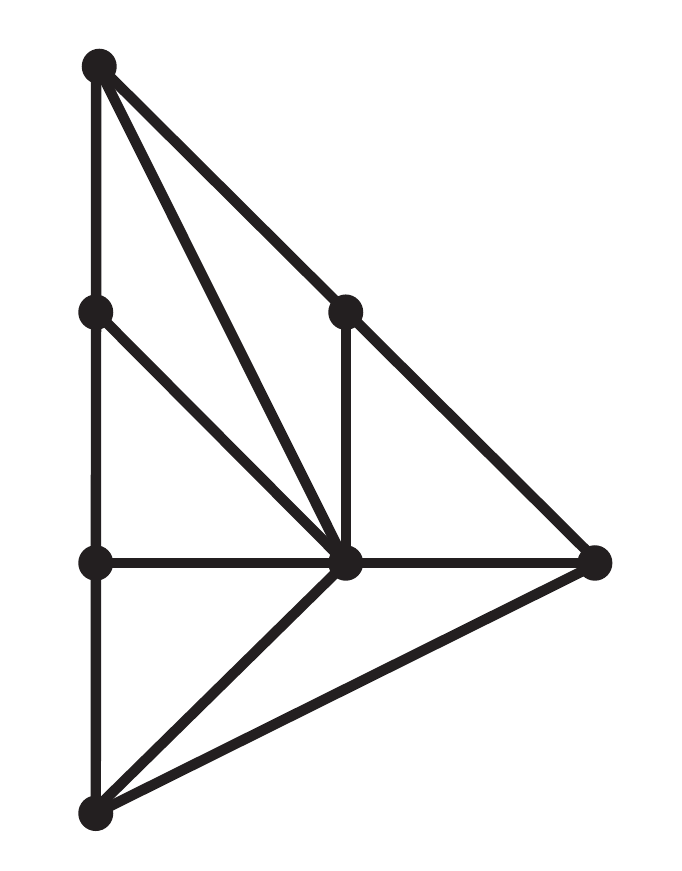} 
  \end{minipage}
  \begin{minipage}[b]{0.45\linewidth}
    \centering
    \includegraphics[height=4cm]{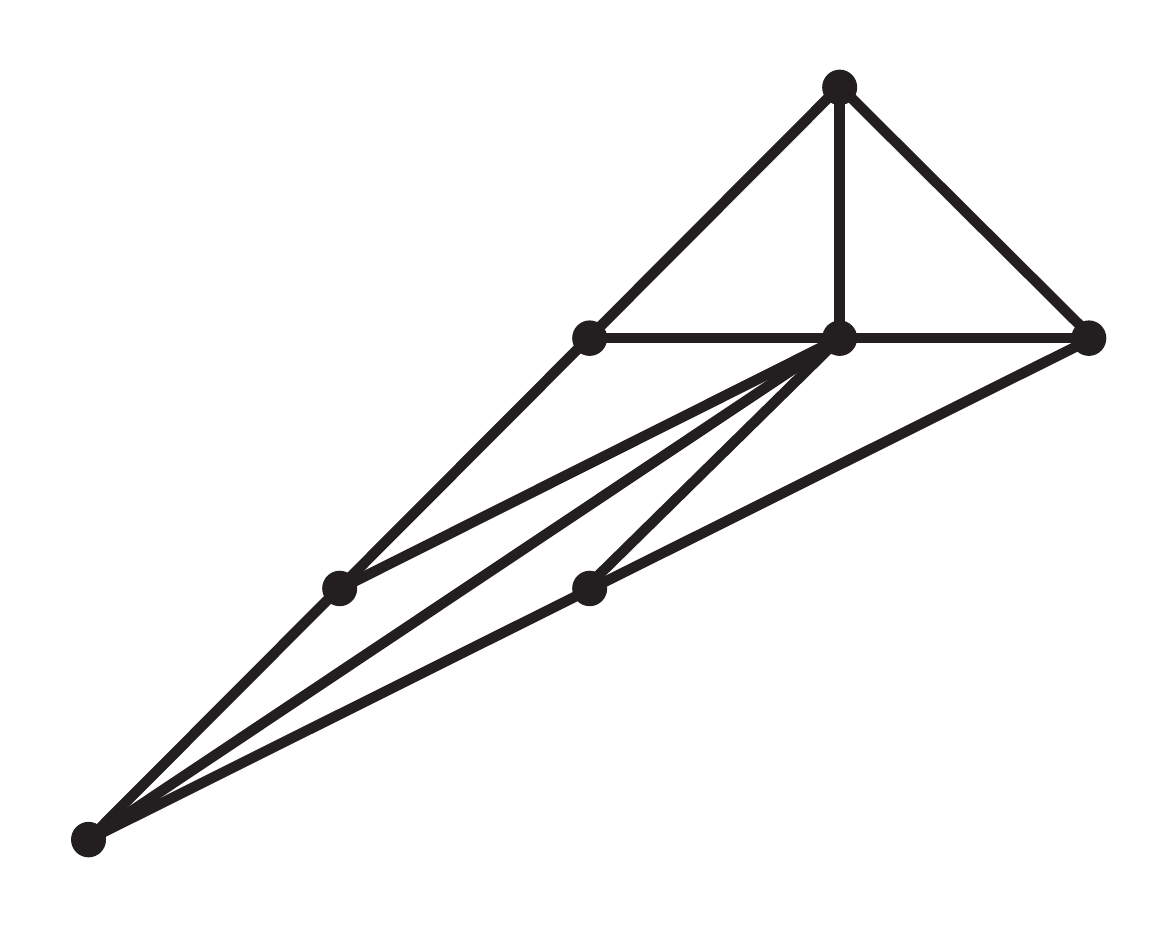}
  \end{minipage} 
\end{center}
  \caption{Two equivalent toric diagrams for the mass deformed $E_8$ del Pezzo geometry.}
  \label{fig:E8}
\end{figure}

A toric diagram of the mass deformed $E_8$ del Pezzo geometry with three non-vanishing parameters 
is given by the left of Figure~\ref{fig:E8}.
The mirror curve of  this geometry thus takes the form%
\footnote{We follow the convention of \cite{HKP} to assign three mass parameters.}
\be
\re^{x}+\re^{-x-y}+\re^{-x+2y}+M_1 \re^{y}+M_2 \re^{-x+y}+M_3 \re^{-x}=\cE'.
\label{eq:mirror-E8-1}
\ee
Changing the variable $y \to -x-y$, we also get
\be
\re^{x}+\re^{y}+\re^{-3x-2y}+M_1 \re^{-x-y}+M_2 \re^{-2x-y}+M_3 \re^{-x}=\cE'.
\label{eq:mirror-E8-2}
\ee
This mirror curve is also obtained by considering the diagram shown in the right of Figure~\ref{fig:E8}.
As in local $\cB_3$, one can reduce this curve to the Weierstrass form \eqref{eq:elliptic-curve} by using Nagell's algorithm.
The coefficients $g_2$ and $g_3$ are now given by (see \cite{HKP} for instance)
\be
\ba
g_2&=\frac{1}{12{z'}^4}[ 1-8M_3 {z'}^2-24M_1{z'}^3 +16(M_3^2-3M_2){z'}^4], \\
g_3&=\frac{1}{216{z'}^6}[ 1-12M_3 {z'}^2-36M_1 {z'}^3
+24(3M_2-2M_3^2){z'}^4
+144M_1 M_3 {z'}^5 \\
&\quad+8(-108+27M_1^2+36M_2 M_3-8M_3^3){z'}^6],
\ea
\ee
where $z'=1/\cE'$.
It turns out that if the parameters are set to be
\be
M_1=\frac{1+m_1 m_2 m_3}{(m_1 m_2 m_3)^{1/2}},\quad M_2=\frac{m_1 m_2 +m_2 m_3+ m_3 m_1}{(m_1 m_2 m_3)^{2/3}},\quad M_3=\frac{m_1+m_2+m_3}{(m_1 m_2 m_3)^{1/3}},
\label{eq:B3-E8-masses}
\ee
and
\be
\cE'=\frac{\cE}{(m_1 m_2 m_3)^{1/6}}
\label{eq:B3-E8-E}
\ee
then the coefficients $g_2$ and $g_3$ are identical to those for local $\cB_3$.
The agreement of the Weierstrass forms implies that the two systems are equivalent as an eigenvalue problem.
We conclude that the two quantum systems associated with \eqref{eq:mirror-B3} and with \eqref{eq:mirror-E8-1} or \eqref{eq:mirror-E8-2}
should give the same eigenvalues if the mass parameters are identified as \eqref{eq:B3-E8-masses}
and \eqref{eq:B3-E8-E}.
In particular, the most symmetric mass choice $(m_1,m_2,m_3)=(1,1,1)$ corresponds to $(M_1, M_2, M_3)=(2,3,3)$.
In subsection~\ref{subsec:test}, we observed that the lowest eigenvalues for local $\cB_3$ with $(m_1,m_2,m_3)=(1,1,1)$
indeed coincides with that for the mass deformed $E_8$ del Pezzo geometry with $(M_1, M_2, M_3)=(2,3,3)$,
computed in \cite{GKMR}.
This correspondence is naturally explained by using the so-called Hanany--Witten transition of $(p,q)$ 5-brane webs.%
\footnote{We are grateful to Futoshi Yagi for explaining it.}
We will report it in detail somewhere else.


\begin{thebibliography}{99}

\bibitem{ADKMV} 
  M.~Aganagic, R.~Dijkgraaf, A.~Klemm, M.~Marino and C.~Vafa,
  ``Topological strings and integrable hierarchies,''
  Commun.\ Math.\ Phys.\  {\bf 261}, 451 (2006)
  [hep-th/0312085].

\bibitem{ACDKV} 
  M.~Aganagic, M.~C.~N.~Cheng, R.~Dijkgraaf, D.~Krefl and C.~Vafa,
  ``Quantum Geometry of Refined Topological Strings,''
  JHEP {\bf 1211}, 019 (2012)
  [arXiv:1105.0630 [hep-th]].


\bibitem{NS} 
  N.~A.~Nekrasov and S.~L.~Shatashvili,
  ``Quantization of Integrable Systems and Four Dimensional Gauge Theories,''
  arXiv:0908.4052 [hep-th].

\bibitem{GS} 
  S.~Gukov and P.~Sulkowski,
  ``A-polynomial, B-model, and Quantization,''
  JHEP {\bf 1202}, 070 (2012)
  [arXiv:1108.0002 [hep-th]].

\bibitem{GHM1} 
  A.~Grassi, Y.~Hatsuda and M.~Marino,
  ``Topological Strings from Quantum Mechanics,''
  Annales Henri Poincare {\bf 17}, no. 11, 3177 (2016)
  [arXiv:1410.3382 [hep-th]].

\bibitem{WZH} 
  X.~Wang, G.~Zhang and M.~x.~Huang,
  ``New Exact Quantization Condition for Toric Calabi-Yau Geometries,''
  Phys.\ Rev.\ Lett.\  {\bf 115}, 121601 (2015)
  [arXiv:1505.05360 [hep-th]].

\bibitem{HM} 
  Y.~Hatsuda and M.~Marino,
  ``Exact quantization conditions for the relativistic Toda lattice,''
  JHEP {\bf 1605}, 133 (2016)
  [arXiv:1511.02860 [hep-th]].

\bibitem{FHM} 
  S.~Franco, Y.~Hatsuda and M.~Marino,
  ``Exact quantization conditions for cluster integrable systems,''
  J.\ Stat.\ Mech.\  {\bf 1606}, no. 6, 063107 (2016)
  [arXiv:1512.03061 [hep-th]].

\bibitem{MZ} 
  M.~Marino and S.~Zakany,
  ``Matrix models from operators and topological strings,''
  Annales Henri Poincare {\bf 17}, no. 5, 1075 (2016)
  [arXiv:1502.02958 [hep-th]].

\bibitem{KMZ} 
  R.~Kashaev, M.~Marino and S.~Zakany,
  ``Matrix models from operators and topological strings, 2,''
  Annales Henri Poincare {\bf 17}, no. 10, 2741 (2016)
  [arXiv:1505.02243 [hep-th]].



\bibitem{GKMR} 
  J.~Gu, A.~Klemm, M.~Marino and J.~Reuter,
  ``Exact solutions to quantum spectral curves by topological string theory,''
  JHEP {\bf 1510}, 025 (2015)
  [arXiv:1506.09176 [hep-th]].

\bibitem{CGM} 
  S.~Codesido, A.~Grassi and M.~Marino,
  ``Spectral Theory and Mirror Curves of Higher Genus,''
  arXiv:1507.02096 [hep-th].

\bibitem{BGT} 
  G.~Bonelli, A.~Grassi and A.~Tanzini,
  ``Seiberg-Witten theory as a Fermi gas,''
  arXiv:1603.01174 [hep-th].

\bibitem{Grassi} 
  A.~Grassi,
  ``Spectral determinants and quantum theta functions,''
  J.\ Phys.\ A {\bf 49}, no. 50, 505401 (2016)
  [arXiv:1604.06786 [hep-th]].

\bibitem{SWH} 
  K.~Sun, X.~Wang and M.~x.~Huang,
  ``Exact Quantization Conditions, Toric Calabi-Yau and Nonperturbative Topological String,''
  arXiv:1606.07330 [hep-th].

\bibitem{CGuM} 
  S.~Codesido, J.~Gu and M.~Marino,
  ``Operators and higher genus mirror curves,''
  arXiv:1609.00708 [hep-th].

\bibitem{MZ2} 
  M.~Marino and S.~Zakany,
  ``Exact eigenfunctions and the open topological string,''
  arXiv:1606.05297 [hep-th].

\bibitem{CSMS} 
  R.~Couso-Santamaria, M.~Marino and R.~Schiappa,
  ``Resurgence Matches Quantization,''
  arXiv:1610.06782 [hep-th].

\bibitem{HKT} 
  Y.~Hatsuda, H.~Katsura and Y.~Tachikawa,
  ``Hofstadter's butterfly in quantum geometry,''
  New J.\ Phys.\  {\bf 18}, no. 10, 103023 (2016)
  [arXiv:1606.01894 [hep-th]].



\bibitem{Hof}  
D.~R.~Hofstadter,
``Energy levels and wave functions of Bloch electrons in rational and irrational magnetic fields,''
Phys. Rev. B {\bf 14} (1976) 2239.  

\bibitem{KM} 
  R.~Kashaev and M.~Marino,
  ``Operators from mirror curves and the quantum dilogarithm,''
  Commun.\ Math.\ Phys.\  {\bf 346}, no. 3, 967 (2016)
  [arXiv:1501.01014 [hep-th]].


\bibitem{Krefl3} 
  D.~Krefl,
  ``Non-perturbative Quantum Geometry III,''
  JHEP {\bf 1608}, 020 (2016)
  [arXiv:1605.00182 [hep-th]].

\bibitem{CW}
F.~H.~Claro and G.~H.~Wannier,
``Magnetic subband structure of electrons in hexagonal lattices,''
Phys. Rev. B {\bf 19} (1979) 6068.

\bibitem{WZ}
P.~B. Wiegmann and A.~V. Zabrodin, ``Quantum group and magnetic translations
  Bethe ansatz for the Asbel-Hofstadter problem,''
  Nucl. Phys. B {\bf 422} (1994) 495--514
  [cond-mat/9312088].

\bibitem{FK}
L.~D. Faddeev and R.~M. Kashaev, ``Generalized Bethe ansatz equations for
  Hofstadter problem,'' 
  Commun. Math. Phys. {\bf 169} (1995) 181--192
  [hep-th/9312133].

\bibitem{HW} 
  M.~x.~Huang and X.~f.~Wang,
  ``Topological Strings and Quantum Spectral Problems,''
  JHEP {\bf 1409}, 150 (2014)
  [arXiv:1406.6178 [hep-th]].

\bibitem{HKP} 
  M.~X.~Huang, A.~Klemm and M.~Poretschkin,
  ``Refined stable pair invariants for E-, M- and $[p, q]$-strings,''
  JHEP {\bf 1311}, 112 (2013)
  [arXiv:1308.0619 [hep-th]].

\bibitem{Dunham}
J.~Dunham, ``The Wentzel-Brillouin-Kramers method of solving the wave equation,'' Phys. Rev. 41
(1932) 713-720.


\bibitem{MiMo}
   A.~Mironov and A.~Morozov, ``Nekrasov Functions and Exact Bohr-Sommerfeld Integrals,''
  JHEP {\bf 1004} (2010) 040
  [arXiv:0910.5670 [hep-th]].


\bibitem{AKMV} 
 M.~Aganagic, A.~Klemm, M.~Mari\~no and C.~Vafa, ``The Topological Vertex,''
  Commun.\ Math.\ Phys.\  {\bf 254}, 425 (2005)
  [hep-th/0305132].
  
  
\bibitem{IKV}
  A.~Iqbal, C.~Kozcaz and C.~Vafa, ``The Refined Topological Vertex,''
  JHEP {\bf 0910}, 069 (2009)
  [hep-th/0701156].


\bibitem{KT} 
  K.~K.~Kozlowski and J.~Teschner,
  ``TBA for the Toda chain,''
  arXiv:1006.2906 [math-ph].

\bibitem{Hatsuda-resurgence}
  Y.~Hatsuda, ``Resummation Problems and Nonperturbative Corrections,'' talk at 
  Resurgence in Gauge and String Theories 2016, Lisbon, Portugal, July 18-22, 2016.



\bibitem{KaMa} 
  J.~Kallen and M.~Marino,
  ``Instanton effects and quantum spectral curves,''
  Annales Henri Poincare {\bf 17}, no. 5, 1037 (2016)
  [arXiv:1308.6485 [hep-th]].

\bibitem{HMO2} 
  Y.~Hatsuda, S.~Moriyama and K.~Okuyama,
  ``Instanton Effects in ABJM Theory from Fermi Gas Approach,''
  JHEP {\bf 1301}, 158 (2013)
  [arXiv:1211.1251 [hep-th]].

\bibitem{HMMO} 
  Y.~Hatsuda, M.~Marino, S.~Moriyama and K.~Okuyama,
  ``Non-perturbative effects and the refined topological string,''
  JHEP {\bf 1409}, 168 (2014)
  [arXiv:1306.1734 [hep-th]].

\bibitem{HKRS} 
  M.~x.~Huang, A.~Klemm, J.~Reuter and M.~Schiereck,
  ``Quantum geometry of del Pezzo surfaces in the Nekrasov-Shatashvili limit,''
  JHEP {\bf 1502}, 031 (2015)
  [arXiv:1401.4723 [hep-th]].

\bibitem{Hatsuda-EQC} 
  Y.~Hatsuda,
  ``Comments on Exact Quantization Conditions and Non-Perturbative Topological Strings,''
  arXiv:1507.04799 [hep-th].

\bibitem{KLS} 
  S.~Kharchev, D.~Lebedev and M.~Semenov-Tian-Shansky,
  ``Unitary
  Representations of $\mathcal{U}_q(SL(2, \mathbb{R}))$, the Modular Double,
  and the Multiparticle $q$-Deformed Toda Chains,''
  Commun.\ Math.\ Phys.\  {\bf 225}, 573 (2002)
  [hep-th/0102180].

\bibitem{Sciarappa} 
  A.~Sciarappa,
  ``Bethe/Gauge correspondence in odd dimension: modular double, non-perturbative corrections and open topological strings,''
  JHEP {\bf 1610}, 014 (2016)
  [arXiv:1606.01000 [hep-th]].

\bibitem{Faddeev1}
L.~D. Faddeev, ``Discrete Heisenberg-Weyl group and modular group,''
  Lett. Math. Phys.
  {\bf 34} (1995) 249--254
[:hep-th/9504111].

\bibitem{Faddeev2} 
  L.~D.~Faddeev,
  ``Modular double of quantum group,''
  Math.\ Phys.\ Stud.\  {\bf 21}, 149 (2000)
  [math/9912078 [math-qa]].

\bibitem{HK}
Y.~Hatsugai and M.~Kohmoto,
``Energy spectrum and the quantum Hall effect on the square lattice with next-nearest-neighbor hopping,''
Phys. Rev. \textbf{B42}(1990)8282.


\bibitem{Sugimoto} 
  Y.~Sugimoto,
  ``Geometric transition in Non-perturbative Topological string,''
  arXiv:1607.01534 [hep-th].


\bibitem{GG} 
  A.~Grassi and J.~Gu,
  ``BPS relations from spectral problems and blowup equations,''
  arXiv:1609.05914 [hep-th].












\end{thebibliography}
\end{document}